\documentclass[12pt,preprint]{aastex}
\def\ltsima{$\; \buildrel < \over \simlt \;$}
\def\simlt{\lower.5ex\hbox{\ltsima}}   

\def\simgt{\lower.5ex\hbox{>sima}}

\begin{document}

\title{The evolution of gravitationally unstable protoplanetary
disks: fragmentation and possible giant planet formation}

\author{Lucio Mayer $^1$, Thomas Quinn $^2$, James Wadsley$^3$, \& 
Joachim
Stadel$^1$}
 
\affil{$^1$Institute of Theoretical Physics, University of Z\"urich, Winterthurerstrasse 190, 8057, Zurich, Switzerland, lucio@physik.unizh.ch} 
\affil{$^2$Department of Astronomy, University of Washington,
Seattle, WA 98195, USA, trq@astro.washington.edu}
\affil{$^3$Department of Physics \& Astronomy, McMaster University, 1280 Main St. West
, Hamilton ON L8S 4M1 Canada,wadsley@physics.mcmaster.ca}

\begin{abstract} We carry out a large set of very high resolution, 
three dimensional smoothed particle hydrodynamics (SPH) 
simulations describing the evolution 
of gravitationally unstable gaseous protoplanetary disks. We consider a broad range of initial disk parameters. Disk masses out to 20 AU range 
from 0.075 to
0.125 $M_{\odot}$, roughly consistent with the high-end of the mass distribution inferred for disks 
around T Tauri stars.Minimum outer temperatures range from 30 to 100 K, as expected from studies of
the early protosolar nebula and suggested by the modeling of protoplanetary
disks spectra. The mass of the central
star is also varied although it is usually assumed equal to that of the Sun. Overall the
initial disks span minimum $Q$ parameters between 0.8 and 2, 
with most models being
around $\sim 1.4$. The disks are evolved
assuming either a locally isothermal equation of state or an 
adiabatic equation of state with
varying $\gamma$. Heating by (artificial) viscosity and shocks is  
included when the adiabatic equation of state is used.
When overdensities above a specific threshold appear as a result of 
gravitational instability in a locally isothermal calculation, 
the equation of state is switched to adiabatic to account for the increased
optical depth. We show that when a disk has a minimum $Q$ parameter less than 1.4 strong trailing spiral
instabilities, typically three or four armed modes, 
form and grow until fragmentation occurs along the arms after about 5 mean disk orbital times.
The resulting clumps
contract quickly to densities several orders of magnitude higher than the initial disk
density, and the densest of them survive even under adiabatic conditions.
These clumps are stable to tidal disruption and merge quickly, leaving 2-3 protoplanets on fairly eccentric orbits (the mean eccentricity being around $0.2$) 
after $\sim 10^3$ years.
Fragmentation is not strongly dependent on whether the disk starts from a
marginally unstable state or gradually achieves it; we show that if the disk 
is allowed to grow in mass from a very light, very stable state over tens of 
orbital times it still fragments at roughly the same mass and temperature 
as in the standard disk models.
We show that the first stages of the instability, until the appearance of
the overdensities, can be understood in terms of the maximum unstable Toomre 
wavelength and the local Jeans length.
A high mass and force resolution are needed to correctly resolve both 
scales and follow the
fragmentation process appropriately. Varying disk mass and temperature 
affects such physical 
scales and hence the typical masses of the protoplanets that form. 
Objects smaller than Saturn or a couple of times bigger than Jupiter can both
be produced by fragmentation. Their final masses
will then depend on the subsequent interactions and mergers with other clumps
and on the accretion of disk material.
The accretion rate depends on the disk thermodynamics and is negligible 
with adiabatic conditions.
After $\sim 10^3$ years the masses range from  just below $1 M_{Jup}$ to more 
than $7 M_{Jup}$, well in agreement with those of detected extrasolar planets.

\end{abstract}


\section{Introduction}

The rapid formation of gas giant planets by gravitational instabilities in a protoplanetary disk (Kuiper 1951;Cameron 1978; Boss 1997)
is an appealing alternative to the conventional scenario of
accretion of gas onto pre-existing large rocky cores formed by accumulation
of planetesimals (Wetherill 1990).
The latter scenario seems to require timescales well in 
excess of disk survival times in dense, highly irradiated environments, 
like the Orion nebula, where most of the stars in our 
galaxies are born (Throop et al. 2001),
making giant planet formation a rare occurrence. Protoplanetary disks in
lower density environments have lifetimes at most marginally consistent 
with the few millions of years required to form a Jupiter sized planet in 
the core-accretion scenario at several AU from their star 
(Pollack et al. 1996, Hubickyj, Bodenheimer \& Lissauer, 2002). However, even
in the most favourable scenario it is hard to imagine how planets 
with masses as large as several Jupiter masses, like many of the observed
extrasolar planets (Marcy et al. 2000; Mayor et al. 2001) might also 
be produced in only a few million years.
The problem is not simply that the planet would migrate inward faster than
it can accrete enough mass (Nelson \& Papalaloizou 2000; Bate et al. 2003).
Indeed inward type II migration 
might be stopped or reversed due to either corotational and Lindblad torques
once more realistic disks with profiles that are not simple
power laws  are considered (Masset \& Papaloizou 2003; Artimowicz \& Peplinski, 
in preparation), but the mass doubling time of a 
Jupiter-sized planet after a gap has been opened is of order of  1 Myr,
even in significantly viscous disks (Artimowicz et al. 1998) ---
the disk might be dissipated before the planet can grow substantially
further. 
In addition, the evidence on inner rocky cores
within the Solar System giants, an inevitable prediction of the core
accretion mechanism, is weakening --- Jupiter might not have a solid
core at all (Guillot 1999a,b). Even for the transiting extrasolar 
giant HD209458b, where
the planetary radius and mass are known (Charbonneau et al. 2000), models
of the planet's interior are generally consistent with the absence of a core 
(e.g. Guillot \& Showman 2002).
The overall amount of metals in 
Jupiter and Saturn is significantly higher than solar, but this does not
necessarily reflect the initial metal content of the planets (Boss 1998).
In any case, current models of the core-accretion mechanism
need a surface density of solids 3-4 times in excess of the minimum
solar nebula model (Weidenschilling 1977) for the rocky cores of
giant planets to form before 10 million years (Lissauer 1993); if augmented
by 99\% times more mass in molecular gas, such a protosolar nebula
model will indeed be marginally unstable to gravitational instabilities,
which will then become the prevailing formation mechanism because it
takes as little as a thousand years (Boss 1997,2000, 2001).
On the other hand, gravitational instabilities in a protoplanetary
disk are hard to treat correctly due to both various numerical pitfalls
that can arise in the simulations and the difficulty of accounting properly
for all the cooling and heating mechanisms present in real protoplanetary disks
(Pickett et al. 1998, 2000 a,b, 2003). 
Due to the complexity of the problem, the natural first step is 
to adopt a simple thermodynamical description of the disk and
use a very high resolution to probe in great detail the highly nonlinear 
dynamics associated with gravitational instability.
In Mayer et al. (2002) we showed that, under
certain conditions, a system of gas giants can arise whose properties 
are reminescent of those of known extrasolar planetary systems. 
Essential to this result was the ability to achieve
very high spatial and mass resolution thanks to the fast parallel Tree+SPH
code GASOLINE (Wadsley, Stadel \& Quinn 2003). 
Here we describe the results of a much larger suite of simulations, exploring 
a wide parameter space in terms of both disk structural properties and
thermodynamics, as well as addressing in detail the various numerical
aspects of the calculations and how these can affect the final outcome
of the gravitational instability. We will also discuss the reliability of the
initial conditions used in the simulations and the structural properties
of the protoplanets formed in some of the runs. In a forthcoming paper we 
will discuss the effects of both irreversible heating and radiative 
cooling in high resolution disk simulations.

\section{Initial Conditions}

The self-gravitating disk models are a collection of 200,000 or 1 million 
SPH particles rotating around a central star.
Disks have masses ranging between $0.075$ and  $0.125 M_{\odot}$,
comparable to the most massive among T Tauri disks (Beckwith et al. 1990;
Dutrey et al. 1996) and 3-4 times more massive than the minimum mass solar
nebula (Weidenschilling 1977). However, these masses are still lower
than those expected in the early stages of the 
formation of the star-disk system from the infalling molecular cloud
material, as suggested by hydrodynamical simulations (Pickett et al. 1998, 2000a).
Initially disks extend from 4 to 20 AU and have a surface density profile 
$\Sigma(r) \sim r^{-1.5}$ with an exponential cut-off at both the inner 
and outer edge.
We do not apply any boundary conditions, a difference with most previous 
works on the subject (e.g. Boss 1998, 2002; Nelson et al. 1998, 2000), so 
disks are free to expand and, at the same time, material can accrete onto 
the central star due to transfer of angular momentum resulting from artificial
viscosity and eventual non-axisymmetric instabilities.
The central star is a softened point mass; its typical mass, $M_s$, 
is equal to $1 M_{\odot}$ (see Table 1), and it can wobble in response
to the time-dependent disk potential; its
softening is 2 AU, so that the innermost annulus of material effectively feels
a point-mass potential --- we use a spline kernel softening (Hernquist \& Katz
1989) for both the
central star and gas particles, hence the potential is keplerian
at two softening scale lengths.
The keplerian velocity of gas particles is corrected
to account for the effect of pressure (gas particles thus move
at slightly sub-keplerian speeds). The initial vertical density structure
of the disks is imposed by the hydrostatic equilibrium for an assumed 
temperature profile $T(r)$.
The disks are not seeded with non-axisymmetric perturbations as 
normally done in grid-based simulations (e.g. Boss 1998) --- Poisson noise is 
indeed present in an SPH simulation even at fairly high resolution 
(at a level of 0.1\% in our 1 million particles runs, comparable to 
the seeds in Boss' models) due to the discrete representation of the system.

The softening of gas particles, $\varepsilon_s$, is constant with time 
in our runs, while 
in some of the previous SPH simulations of the disk instability it 
was evolving (typically decreasing) as the smoothing length
(e.g. Nelson et al. 1998, 2000).
Our implementation is preferred in most astrophysical problems 
as it avoids the unphysical situation of
having particles with varying gravitational potential energy. 
A good choice for the softening is one that allows a gravitational 
force resolution initially close to the resolution of pressure forces, hence 
$\varepsilon_s \sim h$, where
$h$ is the SPH smoothing length calculated over 32 neighbors (Bate \& Burkert
1997). In the latter case the
softening scales with the mass of the gas particles, $m_g$,  
as $m_g^{1/3}$, because so
does the SPH smoothing length; as a result,
we will use smaller softenings in simulations with higher
mass resolution. 
The latter type of scaling between softening and mass is widely adopted and
has been repeatedly shown to be reliable in hydrodynamical
simulations of structure formation (e.g. Hernquist \& Katz 1989; 
Thacker et al. 2000).
As explained in Bate \& Burkert (1997), however, the gravitational
softening must be small enough in order to resolve the local Jeans mass 
even in high density regimes if collapse or fragmentation is the aim
of our calculation. 
The Jeans mass is bigger than the entire system in our disks
initially, but it will eventually decrease if the local density increases
due to gravitational instabilities, and the SPH smoothing length will also
decrease correspondingly. Therefore, in order to leave room for better
force resolution in the later stages of the simulations we choose 
$\varepsilon_s \sim 0.5 h$ (where $h$ is the initial smoothing length)
in the majority of our simulations. 
However, in a few runs we explored the effect of choosing a bigger or
smaller softening, and the results will be discussed in section 3.3.
Gravitational forces between distant particles are calculated using
the hierarchical tree method with opening angle parameter $\theta=0.7$
(Barnes \& Hut 1986). In particular, GASOLINE uses a binary tree
and multipole expansions are carried out up to the hexadecapole order
(see Stadel 2001 for details). The code uses multistepping to advance
particles in an efficient way with a leapfrog integrator (see Wadsley, Stadel \& Quinn 2003).

In most of the simulations the disk is initialized with a mass and temperature 
profile so as to obtain a desired minimum Toomre $Q$ parameter in the outer, 
colder part of the disk, similar to what was done in Boss (2001, 2002a,b). 
We recall that $Q=\Omega v_s/\pi G \Sigma$ for a thin disk in keplerian 
rotation (Toomre 1964), where $\Omega$ is the angular velocity, $\Sigma$ 
is the disk surface density, $v_s$ is the sound speed (which is proportional
to $\sqrt T$, where $T$ is the temperature) and $G$  is the
gravitational constant. Figure 1 shows examples of Q profiles in
some of our models.
We also performed a run in which the disk approaches a given 
minimum Toomre parameter, $Q_{min}$, from a very stable state (high
$Q$), starting from a very low initial mass (run DISLgr in Table 1)
which is grwon over time by an order of magnitude. 
The latter simulation
allows us to test the dependence of our results on the way the 
standard initial conditions are set up. 
The shape of the temperature profile is the same for each model (Figure 2) 
and is similar to that used  by  Boss (1998,2001); 
temperature depends only on radius, thus there is no difference between
midplane and an atmosphere.
Between 5 and 10 AU
the temperature goes $\sim r^{-1/2}$, which resembles the slope obtained if viscous accretion onto the central star is the key driver of disk 
evolution (Boss 1993). Between 4 and 5 AU the temperature profile rises more steeply, being
partially determined by irradiation from the central stellar source in agreement with the 3D  radiative transfer calculations of Boss (1996), while it 
smoothly flattens out for $R > 10$ AU and reaches a constant minimum
temperature (an exponential cut-off is used). 
The minimum temperature ranges between
$35$ and $100$ K (typically is around $50$ K);
it is implicitly assumed that the disk temperature is related 
to the temperature of the embedding molecular cloud
core from which the disk would be accreting material (Boss 1996).
Note that, at least for the protosolar nebula, $50$ K is probably a 
conservative upper limit for the characteristic temperature
at $R > 10$ AU based on the chemical composition of comets in the Solar System
(temperatures as low as $20$ K are suggested in the recent study by
 Kawakita et 
al. 2001). Outer temperatures between $30$ and $70$ K are found also for several T Tauri
disks by modeling their spectral energy distribution assuming a mixture of gas
and dust and including radiative transfer (D'Alessio et al. 2001).

In addition to the $Q$ parameter, another important measure of the  
susceptibility of a disk to gravitational instabilities is provided by the $X_m$ parameter, $X_m={\Omega}^2 R/2 m \pi G \Sigma$, $m$ being the order
of the unstable mode and $R$ being the disk radius. An extensive literature of
numerical experiments conducted for both collisionless and gaseous disks
has shown that $X_m$, coupled with $Q$, provides a good measure of
 the susceptibility of the
disk to swing amplification of a given mode. In swing amplification a
leading wave is amplified into a higher amplitude trailing wave, and 
if the latter can be turned back into a leading wave a feedback
loop is initiated that can produce a disturbance whose magnitude
is orders of magnitude greater than that of the initial wave
(Binney \& Tremaine 1987).                                            
In Figures 1 and 3 we show the $Q$ profile
and $X_m$ profiles ($m=2,3$) for two of our disk models; other models
(see Table 1)  differ only
in the value of  $T$ or $\Sigma$ and hence their profiles
can be easily recalculated 
(they differ only in the normalization, not in the shape).
Strong swing amplification typically requires $X_m < 3$ and $Q < 3$ in
some region of the disk (Binney \& Tremaine 1987); both
conditions are marginally satisfied only at the very edge of the disk  
for $m=2,3,4$ in most of our models (e.g. DISL1 and DISH1, see Figure 3), 
while they are definitely satisfied in model DISL4 at $R > 15$ AU (see Table 1;
this model has a central star whose mass, $M_s$, is half of the standard value, and as
$\Omega^2 \sim M_s$, $X_m \sim M_s$, hence $X_m$ is a factor of 2 lower 
for any $m$ compared to other disk models having the same mass). In section 3.1.1 we will
discuss the role of that the initially sharp outer edge of the disks 
might have in swing amplification.

We performed both locally isothermal (ISO) and adiabatic (ADI) runs. 
The gas is always assumed to be purely in the form of molecular hydrogen, 
hence we assume $\mu=2$ for the molecular weight. In what we call the 
adiabatic runs we solve a thermal energy equation which also includes 
heating from artificial viscosity (see below), in particular the 
quadratic term in the latter accounts for irreversible shock heating
(Monaghan \& Gingold 1983). 
The equation of state has the
form $P=(\gamma - 1)\rho u$, where $P$ is the pressure, $\rho$ is the density
and $u$ is the specific internal energy of the gas and $\gamma$ is the
ratio between the specific heats. 
We assume $\gamma=1.4$ in most of the simulations(appropriate for molecular hydrogen with roto-vibrational transitions) but,
more in general, we vary it in the range $1-1.4$. 
The locally isothermal equation of state stands on the assumption that 
any heating is instantaneously radiated away. The latter is 
simply written as $P=\rho u$, where $u$ is now constant (therefore
no thermal energy equation is solved in this case) and is given by 
$u=k_B T / \mu$, where $T$ is the temperature and $k_B$ is the Boltzmann
constant. 
In our simulations the gas is isothermal in a Lagrangian sense, i.e. the 
thermal energy of a given particle is assigned based on its initial 
distance from the star and does not vary, irrespective of its subsequent
motion through the disk. 
We note that,
although some radial mixing is expected to occur because of developing
non-axisymmetric instabilities, the initial temperature of the disk 
is constant by construction throughout most of the outer disk, where the 
strongest instabilities should develop; we verified that radial 
motions of particles do not have any 
significant impact on the outer disk temperature profile, thereby 
the hypothesis of local isothermality is self-consistent in the regions of 
interest (Figure 2).
Values of $\gamma$ smaller than the canonical $1.4$ yield smaller pressure and 
are expected to produce results closer to those of the locally isothermal
runs.

All our simulations include artificial viscosity in the standard Monaghan
formulation (Monaghan \& Gingold 1983)
plus the Balsara correction term (Balsara 1995)
 to reduce unwanted viscosity in purely
shearing flows (see Wadsley, Stadel \& Quinn 2003 for details). Artificial viscosity appears in both the momentum and the thermal energy equation.  
In SPH codes
artificial viscosity is introduced for a number of reasons, primarily
to avoid particle interpenetration and reduce post-shock oscillations in
high Mach number flows. The magnitude of the viscosity
terms becomes smaller with decreasing smoothing length, hence with increasing
resolution.  In most of the
runs the linear and quadratic coefficients of artificial viscosity are set to,
respectively, $\alpha=1$ and $\beta=2$. Although this choice is standard in
three-dimensional SPH calculations (see for example Hernquist \& Katz 1989, Navarro \& Benz 1991, Thacker et al. 2000), it is still borne out of classic tests like the shock tube and the isothermally collapsing gas cloud, and it is not
guaranteed to be optimal in more complex systems like those considered here.
In particular, whereas the quadratic term is needed to properly follow the
shocks that will eventually develop during the gravitational instability, the
linear term mainly damps the velocities of particles, reducing the noise
inherent to the SPH technique.  One worry is that numerical
viscosity, by acting as an effective pressure, might smear out even physical
small-scale features in the velocity field (for example in a region
that is about to collapse due to gravitational instability) and 
generate spurious angular momentum transfers (Thacker et al. 2000). 
Heat generated by artificial viscosity when the thermal energy equation is
solved might also affect the disk evolution.
We investigate how artificial viscosity affects our results in both isothermal and adiabatic calculations by varying the value of the $\alpha$ and $\beta$ 
coefficients (see Table 1 and section 3.4).

\section{Results}

In what follows we describe the results of our large suite of simulations. 
The various setups are indicated in Table 1.
In the same Table we also indicate whether clumps (protoplanets) are formed
or not in a given run.
Disks are typically followed for about 12-15 orbital times, where we
define the characteristic orbital time as that at $R=10$ AU (halfway 
between the center and the edge of the disk), 
$T_{orb} = 2 \pi \sqrt{(GM_s/{R^3})} = 28$ years.
Disks that undergo fragmentation are generally evolved for the shortest
timescale (12 orbital times) because the growing local overdensities 
require increasingly smaller timesteps to be accurately followed, 
with the result of slowing down considerably the computation. 
We use up to $400,000$ timesteps for 
the most expensive calculations. Two simulations (DISL1 and DISgr) were 
carried out for a much longer timescale, about 20 and 30 orbital times, 
respectively, and their results will be described in section 3.5,
together with the structural and orbital properties of the formed protoplanets.

All disks develop trailing spiral instabilities after a few rotations, 
the strength and nature of which depends on the $Q$ and $X_m$ parameters 
and on the equation of state. The disk expands because of the spiral arms, 
which shed angular momentum outwards and mass inwards; as a result its
profile becomes more concentrated with time (Figure 4).
Although initially the minimum $Q$ of the disks, $Q_{min}$ is located at 
their outer boundary, their rapid expansion in response to the spiral 
instabilities 
causes a drop of the surface density in the outer part, and therefore 
$Q_{min}$ shifts further inward, near 13-14 AU (see Figure 1).
The larger pressure gradients developing in adiabatic runs tend to
erase disk substructure created by gravitational instability, while locally
isothermal runs provide the most favourable conditions for the developing
of the instabilities through damping of those same pressure gradients. 
The nature of the spiral pattern, namely which modes are dominant, depends
on the details of disk structure. In general we observe that, for a
given $Q$, disks with lower masses and temperature tend to produce
higher order spiral patterns. This is likely related to swing
amplification being stronger for higher order modes in lighter disks
($X_m \sim {(\Sigma m)}^{-1}$)  
and has been previously observed by other authors 
(e.g. Nelson et al. 1998, 2000;Rice et al. 2003).

\subsection{Locally isothermal runs}

Disks evolved isothermally undergo fragmentation for $Q_{min} \le 1.4$. In
these disks $Q$ drops below unity between 12 and 15 AU after $\sim
200$ years, and shortly after several
clumps appear that become gravitationally bound ($2P + U < 0$, where
$P$ is the pressure and $U$ is the gravitational binding energy) over
a fraction of the orbital period, reaching densities of order $10^{7}
g/cm^3$ in their centers, up to 6 orders of 
magnitudes higher than the initial local density, their further collapse being
limited only by numerical resolution (see section 3.3). When the local
density grows beyond ten times the initial local value the gas should behave 
nearly adiabatically according to the radiative transfer calculations
of Boss (2002) due to the increase in opacity (we recall that our disks
have initial surface densities roughly identical to those in Boss' models) and 
therefore the locally isothermal approximation is no longer valid. 
Clump formation
still proceeds when we  switch to adiabatic conditions once 
the critical density threshold is reached. 
Clumps are fewer in the latter case but several gravitationally
bound ones are still present (Figure 5). We note that
our adiabatic conditions include irreversible heating from artificial 
viscosity --- the temperature near the spiral overdensities indeed rises 
above the isothermal value, reaching $80$ K --- yet this is not enough 
to suppress clump formation at this stage because of the
high density contrast already achieved (see Mayer et al. 2002). For $Q_{min} \sim
1.65$ strong spiral arms are observed, but these saturate at some point
(Laughlin, Korchagin \& Adams, 1997) 
reaching a near stationary pattern after almost 20 orbital times. At even larger values $Q_{min}$, $\sim 1.9$, 
very weak spiral arms form and then saturate. The
evolution of the $Q$ parameter strongly suggests that the threshold
between fragmentation and self-regulation must lie near $Q \sim 1.4$.
Indeed models  whose initial $Q_{min}$ is $1.65$ reach $Q_{min} \sim 1.15$ locally (between 12 and 15 AU), which is only slightly higher than 
the $Q \sim 1$ required
for fragmentation. A simulation with $Q_{min} \sim 1.5$ also did not lead to
fragmentation (see Figure 6); the spiral arms reach a consistent amplitude
after about 200 years, but then weaken and saturate.
Therefore $Q_{min} \sim 1.4$ really seems to mark the threshold for 
fragmentation in our calculations, but
in general such threshold will vary depending on the structural
properties of the disk (e.g. its surface density profile).

As mentioned in section 2, the $Q$ profile is not enough to
characterize the evolution of the different disk models; in addition to
having different types of spiral patterns, disks with same $Q_{min}$ but
different temperatures/masses yield clumps with varying mass. 
In particular, for a given $Q_{min}$ 
lighter and colder disks produce less massive clumps, but the mass
of the clumps does not scale linearly with the mass of the disks.
For
example, model DISH3, which has a mass of $0.085 M_{\odot}$, produces several
protoplanets with masses below a Saturn mass, which eventually grow up to
a Jupiter mass or slightly above that (see below); therefore in this case
protoplanets have masses up to 3-4 times smaller than those arising in
disks only 15\% more massive (e.g. model DISL1). 

The trend can be understood in terms of the dependence of the local Jeans 
mass on disk temperature and mass. Only overdensities whose scale 
is above the Jeans length and below the Toomre
critical wavelength (see Binney \& Tremaine 1987) will be able to grow
and survive.  In particular, any overdensity whose size is 
larger than the Toomre wavelength will be sheared away by differential 
rotation irrespective of disk temperature, and it will not grow 
in the first place unless its mass and size are above
the local Jeans mass or length, respectively. Initially the Jeans length is
comparable to the size of the entire disk, but as the instability proceeds
and the disk midplane density increases it falls down to values below or
around a Jupiter mass depending on the disk model. For disks having the
same $Q_{min}$, the Jeans mass is smaller for lighter (colder) disks. 
In fact, from the definition
of the Jeans mass we have $M_J= {\pi \over 6} \rho{\left({\pi {v_s}^2} \over
 {G \rho}\right)}^{3/2}$ ($\rho$ is the density, $v_s$ the sound speed and $G$
the gravitational constant), hence $M_J \propto T^{3/2}$. But the scaling 
between density and temperature at fixed $Q$ is $\rho \sim T^{1/2}$ so 
ultimately $M_J \sim T^{5/4}$. A difference of a factor of
2 in the temperature (required to compensate a difference of order
$\sqrt 2$ in the disk mass) introduces a difference of a factor $\sim 2.4$ 
in the Jeans mass and hence in the minimum mass expected for condensations.
Figure 7 shows that, indeed, there is a factor of 2 or more variation in the
minimum mass of the clumps at the onset of fragmentation when we compare
disks with different masses and same initial $Q_{min}$.

Is also the maximum size of the overdensities consistent with simple 
theoretical expectations?
According to the tight-winding (WKB) approximation, which is only valid 
for tightly wound (local) perturbations in a differentially rotating thin 
disk, only density perturbations  whose scale is smaller than the 
Toomre wavelength, $\lambda_{crit}=4 {\pi}^{2} G \Sigma / {\kappa}^2$,
can grow. 
The most unstable wavelength is  
$\lambda_{mu} = 0.55 \lambda_{crit}$ for zero-thickness gaseous disks
(Binney \& Tremaine 1987).
For most of our disks, for which $M_d=0.1 M_s$, we have 
$\lambda_{mu} \sim 5$ AU at 
distances  between 12-16 AU, the region where fragmentation occurs, 
and because 
$\lambda_{mu}/R \sim 0.25$, i.e. $\lambda_{mu}/R << 2 \pi$ in the same 
range of 
radii, WKB results should still be valid for axisymmetric waves 
(Binney \& Tremaine 1987). For non-axisymmetric waves, the
condition $X_m >> 1$ (where $X_m$ is the parameter defined in section 2) is
more appropriate, but this is also satisfied throughout most of the disk
in the standard models, up to $m=3$ (but see below on the swing amplification).
We would 
expect clumps to occur at scales of order $\lambda_{mu}$ and always 
below $\lambda_{crit}$. In disks with non-negligible vertical pressure, and
hence finite thickness, both wavelengths
will be somewhat smaller as the disk will have an effective self-gravity
lower than in the truly thin case. Numerical softening, in addition, can
also be thought as providing an artificial pressure on small scales.
Romeo (1992; 1994) has computed correction factors for the 
stability properties of self-gravitating 
stellar and gaseous disks which account for finite thickness and numerical softening. These factors are formulated as reduction factors for the surface density of the disks. In particular, if the wavenumber corresponding to the most unstable wavelength is $k_{mu} = {2 \pi}/\lambda$, and the typical 
disk scale height resulting from pressure is $h_d$, then the correction 
factor for finite disk thickness reads ${(1 + k_{mu}h_d)}^{-1}$; an identical
expression can be obtained for the softening $\varepsilon$, where $\varepsilon$
replaces $h_d$. In most of our simulations $\varepsilon << h_d$, and 
hence we only consider the first reduction factor. Indeed, we
find that  $\lambda_{mu}$ is reduced by more than $60 \%$ once we include 
the latter factor, becoming $\sim 2$ AU, which turns out to be 
the typical size of the banana-shaped overdensities that appear along the spiral arms just {before fragmentation. These overdensities
rapidly fragment into multiple clumps, but during this phase the
system has become so strongly non-linear that any extrapolation of linear
theory is meaningless. However, it is remarkable that while the
system is still mildly non-linear the results of WKB theory are in good
agreement with the simulations.

One ingredient that is not captured by the WKB approximation is swing
amplification of non-axisymmetric modes. Indeed when strong swing 
amplification is expected  it also means that linear theory breaks down
(Binney \& Tremaine 1987). As we mentioned in the
previous section, most of our disk models (e.g. DISL1, DISH1, DISH2) 
are barely in the
regime where swing amplification is expected to be important at $R > 10$ AU. In particular, only for $m=3$ (or even higher order) modes 
the condition $X < 3$, necessary for strong swing amplification (see Binney 
\& Tremaine 1987), becomes satisfied at some point and only quite late 
in the evolution (see Figure 3).
However, model LIS4, that has a lower mass of the central star, clearly
departs from this picture as $X < 3$ for both $m=2$ and $m=3$ modes
already early during evolution (this suggests that WKB results are 
less applicable here).
Indeed the latter model exhibits a mixture of 3 and 2 armed modes visibly
stronger than in the standard models (3 armed spirals are evident but two
of the three arms grow more in amplitude and they begin to 
fragment, similarly to other simulations), as shown by comparing Figure 8
with Figure 6. Clumps appear almost 2 orbital times earlier in this model (the
orbital time at $10$ AU is 40 years), after only
160 years, and the stronger non-axisymmetry produces orbits with 
eccentricities as large  as $e=0.35$, thereby bigger than in the other
models (see Mayer et al. 2002 and section 3.5 of this paper),
which then favour more frequent and violent interactions 
between the clumps.

\subsubsection{The problem of the initial conditions}

How does fragmentation depend on the initial conditions of the disk
simulations? An often heard
argument against fragmentation models states that simulations with disks
having a low $Q$ parameter by construction are unrealistic because
the disk would respond to a developing gravitational instability
by rising its temperature and adjusting its density profile, thus
self-regulating to fairly high $Q$ values
(Laughlin \& Bodenheimer 1994; Laughlin \& Roczyska 1996; Laughlin, Korchagin
\& Adams 1997). The argument was originally developed against
models starting with $Q$ near unity somewhere in the disk but one
can easily imagine that mass redistribution due to milder non-axisymmetric 
instabilities might always drive the disk to values higher than even the critical threshold 
that we claim here, $Q \sim 1.4$. Moreover, in the simulations
presented so far the disk has a sharp edge at $t=0$; although the disk very
quickly (in one orbital time) expands as a result of its own evolution, 
initial reflection of waves at the edge might artificially activate
a feedback loop and sustain swing amplification, leading artificially to 
a faster and stronger growth of the instability.  Motivated by these arguments
we performed a simulation (run DISgr in Table 1) in which the mass of the disk is increased gradually
(over about 600 years, 20 orbital times at $R=10$ AU) by a factor of 10, going
from $M_d = 0.0085$ (well below even the minimum mass solar nebula) to
$M_d=0.085$ (the final mass is the same as in model DISH3). The mass
is increased uniformly in the disk at every step (in practice we grow the
mass of each particle by a constant fixed). The disk is evolved isothermally
in the first phase, and then adiabatically once the usual overdensity threshold
is reached; the initial temperature profile is of the usual form (see section 2), and in particular the
outer minimum temperature is $30$ K. 

The disk stays considerably smooth for a long time (the initial $Q_{min}$ is higher than 10), but as 
$Q$ approaches 2 spiral instabilities start to appear and grow in amplitude
while the mass continues to increase (disk snapshots are shown in Figure 14),
until $Q$ drops near unity (Figure 9) and clump formation occurs. Note that
the disk surface density has not been redistributed 
significantly since fragmentation occurs at roughly the same temperature and
the same mass as those initially assigned to one of the fragmenting 
standard models initialized with a low $Q_{min}$, model DISH3 (see Table 1),
This experiment
shows that mass redistribution when $Q$ is higher than $1.4$ or so 
is not important --- non-axisymmetric torques are still too weak in this
regime and the disk can still undergo clump formation provided that
it remains cold. In addition, in this test the sharp outer edge disappears
several orbital times before the first weak non-axisymmetric pattern
becomes apparent in the disk, yet fragmentation takes place more
or less as in the other runs. We only note that in this simulation
the dominant spiral patterns are of higher order ($m=5$ or higher) than
in any of the other runs; this is in part due to the fact that the
disk enters the regime of non-axisymmetric instability with a mass
lower than in any of the other models, so that swing amplification 
is effective for high order modes only during the first part of the 
evolution (see Figure 9), but we
cannot exclude that which is the dominant spiral pattern also depends on
whether or not initial edge effects are present. 
The outcome of this run is in agreement with another result that points 
to negligible mass redistribution when $Q > 1.5$; the fact that disks 
with $Q_{min} \sim 1.65$ maintain a value of $Q_{min}$ similar to the 
initial one after the strongest
phase of the instability is over (see Figure 1). Of course in reality
the process of mass accretion from the molecular cloud is much more
complicated than depicted here; the disk mass will not increase 
monotonically but will probably reach a maximum and eventually decrease 
as accretion onto the central star or photoevaporation take over accretion
from the molecular cloud (Matsuyama
et al. 2003). In addition, and most importantly, the temperature will not 
remain constant but will change as a result of heating and cooling.

\subsection{Adiabatic runs}

Disks evolved adiabatically since $t=0$ reach fragmentation only when 
starting from very low values of $Q_{min}$, 
as small as $0.8$. Such low values of $Q_{min}$ are obtained with a combination of low temperatures 
and high masses ($M_{disk} = 0.125 M_{\odot}$) (see models marked ``ad''
in Table 1).
With these very low values of the Toomre parameter the disk is locally unstable to axisymmetric perturbations. 
Indeed we see a ring forming in the outer part of the disks, but this is soon 
dissipated as $Q$ rises due to heating by compressions and shocks, and
spiral arms form after 2-3 of orbital times. 
The precise value of $Q_{min}$ needed 
for fragmentation decreases with increasing values of $\gamma$ in the
equation of state.
The results are listed in Table 1 and shown in Figures 10 and 11.
We note that even disks with $Q_{min} \sim 0.8$ undergo
fragmentation only if $\gamma=1.2$, which would happen only if some
cooling is present. Also for $\gamma=1.3$ does fragmentation
occur for  $Q_{min} \sim 0.8$, but the clumps are quickly washed out by the strong 
developing pressure gradients, while for $\gamma=1.4$ overdensities in
the spiral arms are washed out before being able to collapse.
The temperature rises to more than $100$ K along
the spiral arms due to compressional and shock heating.
Therefore, even at these very low
values of $Q$ we do find that fragmentation depends on the
equation of state, while  an early  work by Boss (1998) was
finding fragmentation even for $\gamma=1.4$ with $Q_{min} \sim 1$.
Our tighter limits are likely due to the
inclusion of irreversible shock heating, which was absent in that
as well as in other studies, and is instead important in these 
violently unstable models. Shock heating is evident along the trailing
and leading edges of the spiral arms (see the temperature maps
in Figure 11); indeed these are the locations where the velocities
of the gas in the spiral arms and the mean sound speed differ
the most.

As we mentioned above, these very low $Q$ states are probably unrealistic,and
this is shown by the rapid evolution that the disk undergoes, with $Q_{min}$
rising by a factor of $2$ in less than 5 orbital times. The
clumps that form in run DISad4 are particularly big (see section 3
on how disk mass and temperature influence clump formation) 
growing to more than 10 Jupiter masses after a few mergings. 
Finally, we note that in adiabatic simulations that start from higher
values of $Q_{min}$ (DISLad1,DISLad5,DISLad6 in Table 1), fragmentation
does not take place. The temperature along the spiral arms, on the
other end, only grows from $56$ 
to about $70-80$ K because the non-axisymmetric modes are much weaker than
in the runs with lower $Q_{min}$; this temperature is only $\sim 50$\% 
higher than that
required to maintain $Q \sim 1.4$ and follow the path towards strong
instability (the change in temperature has a negligible dependence on
the magnitude of the artificial viscosity, see Table 1 for the different
cases considered). 
Therefore shock heating is not dramatic  and only little 
cooling would be needed to bring the disk towards the marginally unstable 
regime; on the rise in temperature associated with
shock heating occurs very fast, over less than an orbital time, so
it remains to be seem whether cooling processes can act at comparable
speed (but see Boss 2002a).

\subsection{Mass and force resolution; how they affect clump formation
and evolution}

The physical interpretation of fragmentation 
proposed  in section 3.1 clearly implies that the simulations ought to 
resolve the Jeans and Toomre wavelengths.
The ability to resolve these characteristic physical scale lengths depends on both mass and force resolution in an SPH simulation; in particular, resolution
must be high enough to resolve the smallest among the two scales,
usually the Jeans length. As extensively discussed 
by Bate \& Burkert (1997), the gravitational softening and the SPH  smoothing
length should be comparable and both smaller than the local Jeans length
for the calculations to be trustworthy (see also Nelson 2003).
When the gravitational softening
exceeds the SPH smoothing length any eventual collapse will be slowed down
or halted, while artificial fragmentation might occur in the opposite
situation. Bate \& Burkert also showed that the local Jeans mass should be
always resolved by no less than $2 N_{neigh}$, where $N_{neigh}$ is the
number of neighbors used in the SPH smoothing kernel, $N_{neigh} = 32$ in
our runs. Thanks to the extremely high resolution adopted in this work,
the local Jeans mass is always resolved by several hundred to several 
thousand particles, hence we are orders of magnitude above the minimum
requirements. On the other end, shortly after clump formation the smoothing
length becomes significantly shorter than the gravitational softening inside
the clumps and hence the collapse of the clumps is ultimately slowed down 
once they have shrunk down to a scale comparable to the softening, which
then acts  like an artificial pressure force. This means that in the 
isothermal calculations
the clumps become effectively super-adiabatic near the center, where the 
resolution limit  is reached --- the difference with the
calculations in which the equation of state is switched to adiabatic exists 
only in the very first stage of clump formation, when the
softening is still comparable to the other relevant scale lengths.
The bottom line is that any detailed analysis of the internal structure of 
the clumps must be postponed to future simulations with even higher resolution.
For the moment only mean properties of the clumps, calculated as averages
over the entire systems (which is comfortably larger than the softening)
can and will be discussed (see section 3.5).

A very important result emerging from our set of simulations is that clump formation is enhanced with 
increasing mass resolution for the same initial conditions (with
the softening scaling as ${m_g}^{1/3}$; a larger number of clumps is seen 
at higher 
resolution, although the mass  scale of fragmentation is basically 
unaffected (Figure 12).
Indeed the minimum clump mass depends on the local Jeans mass, which in turn 
is determined by the disk density and temperature, not by resolution (section 3.1).
Force resolution alone also has an effect.
A softening larger or comparable to the 
maximum unstable Toomre wavelength suppresses clump formation even in unstable
disks; in fact a large softening acts as an additional pressure (in
practice it suppresses gravity) at the crucial scale where the perturbation
has the highest amplitude, or, alternatively, one can view it as producing
an effective increase in the local Toomre parameter. Following Romeo (1994),
the minimum allowable softening for our disks should be around $0.37$ AU,
which, of course, is close to the effective $\lambda_{mu}$ calculated
in section 3.1. Actually we find that a softening $\sim 3$ times smaller than
the latter is necessary 
to go beyond the stage of the mildly non-linear regime and enter that of clump
formation (see Table 1).
Indeed the calculation done by Romeo applies to the study of
spiral structure in marginally unstable (galactic) disks and not to the 
strongly unstable regimes that we are investigating here; when $Q$ locally 
drops below $1$ and the strongly non linear regime is reached a higher force
resolution, close to the rapidly dropping Jeans length, should be 
required in order to keep following the dynamical evolution properly. 
Therefore, in a simulation both mass and force resolution must be 
high enough to follow  the fragmentation process in the
disk. 
The maximum and minimum allowable softenings can be accurately determined
only through convergence tests,  the only {\it a priori} prescription
being to balance softening and smoothing length (Bate \& Burkert 1997).
We note, however, that the line dividing stable and unstable disks is
only weakly dependent on softening in our simulations;
disks that are stable with our "standard" choice of parameters remain stable 
even with a softening ten times smaller (for example compare run DISH4
and DISH4b in Table 1). This means that the threshold for stability, 
$Q_{min} \sim 1.4$, is a robust physical result, at least 
under the thermodynamical conditions adopted.

\subsection{Dependence on artificial viscosity}

We find that both the value of $\alpha$ and $\beta$ have some impact on disk
evolution and fragmentation. Whereas disks that do not fragment are found to 
behave so irrespective of the artificial viscosity, disks that fragment do so
more or less severely and on slightly different timescales depending on the
value of these parameters. Here we discuss the results of locally isothermal
runs with varied viscosity parameters (all listed in Table 1), hence we
neglect the artificial heating eventually induced by viscosity. The latter
is present in adiabatic runs, but we already showed that in such
runs disks  never form bound clumps unless we reduce the
value of $\gamma$ significantly below the canonical $1.4$ (see section 3.2),
which makes them rather unsuitable for analyzing the effects of viscous
heating on clump formation. Therefore we decide to
postpone the analysis of artificial viscous heating to a forthcoming paper
in which we implement also radiative heating and cooling in the disks 
(Mayer et al., in preparation).

In general a smaller value of $\beta$ or $\alpha$ 
enhances fragmentation and the opposite happens for larger values. Only $\beta
\ge 3$ can completely suppress fragmentation; however
such high values of $\beta$ are not a good choice for flows with moderate Mach numbers
($\sim 1-1.5$) like those occurring in our simulations (see e.g. Hernquist \& Katz 1989; Thacker et al. 2000) but could mimic the behaviour of disks with a high degree of turbulence (Nelson et al.1998). 
Why is viscosity affecting clumping? In general artificial viscosity
makes the velocity and density fields smoother, which helps to reduce
post-shock oscillations and noise but could in principle suppress 
small-scale physical features in the velocity field of the fluid.
When the collapse of an overdensity begins, particles locally acquire radial 
motions, and hence the radial velocity dispersion will rise; a high viscosity
can damp these radial motions and hence the collapse. 
The fact that the disk velocity dispersion profile is both lower and smoother
with higher artificial viscosity is an indication of such an effect (Figure 13).In particular, disk models that produce several 
bound clumps with the standard values of artificial viscosity (DISL1),
clearly have their localized peaks in the radial dispersion profile
completely smeared out for higher viscosity (DISL1e and DISL1f).
The question arises whether the dependence
on artificial viscosity is strong enough to change the threshold $Q$ for
fragmentation. The answer is negative based on our simulations --- 
disks with $Q$ above the threshold remain stable irrespective of the 
values of $\alpha$ and $\beta$. 
This is highlighted by the comparison between runs DISH2 and DISH2b , in which a disk with 
initial $Q_{min} = 1.65$ is evolved first with the standard
parameters and then with $\alpha=0$ and $\beta=0.5$; while the velocity
dispersion profiles look somewhat different, the dynamical evolution is
substantially identical, and in both runs the disk does not fragment.

\subsection{Long term evolution of disks and protoplanets}

Once the disks enter the fragmentation phase timesteps become extremely
small locally, and the computation becomes extremely demanding. One million
particle runs require almost 400,000 steps up to 350 years of evolution,
80\% of which cover only the last 2-3 orbital times ($\sim 100$ years), after
clumps begin to form. We resort to lower resolution runs (200,000 particles)
to probe the disk evolution over a more extended time (see also Mayer et al. 2002). 
In these runs the thermodynamics switch to adiabatic
as soon as the local density becomes ten times higher than the initial value
(see section 3.1).
The central temperature of the clumps grows up to $300-400$ K rapidly
after formation takes place owing to strong compression while their 
collapse proceeds on their internal dynamical time (of order of days)
; however, the clumps would 
certainly become much hotter in the center if their collapse was not halted at a scale
comparable to the gravitational softening.
Two low-res simulations ($N=200,000$
particles), one for a disk with mass $M=0.1 M_{\odot}$ (same as in run DISL1)
and  one for the growing disk (run DISgr, final mass $M=0.085 M_{\odot}$) 
were run for, respectively, 30 and 20 more orbital times (the reference
orbital time being measured at 10 AU)  
after the onset of the fragmentation. Due to their extremely
high densities protoplanets are never destroyed by the tidal field
of the central star (their tidal radius is more than ten times
larger than their typical size) but are tidally perturbed by and undergo a 
series of mergers with neighboring protoplanets (this phase lasts
about 10 orbital times)  until only 
3 and 2 protoplanets remain in, respectively, model DISL1 (see
also Mayer et al. 2002) and model DISgr
(see last snapshot of Figure 14 for run DISgr).
In both simulations protoplanets have eccentric orbits, with the 
eccentricity, $e$, running between 
0.1 and 0.3 ($e=(R_{apo} - R_{peri})/(R_{apo} + R_{peri})$, where $R_{apo}$
and $R_{peri}$ are, respectively, apocenter and pericenter distance.) 
and end up at mean distances between 3 and 12 AU. These eccentricities
correspond to the mean values found for extrasolar planets (Marcy \& Butler 1998; Marcy et al. 2000).
Larger eccentricities (comparable to the highest measured for extrasolar
planets, $\sim 0.7$) are measured in a run in which the same disk as in run DISL1 is evolved with a locally isothermal equation of state throughout 
the evolution (see below) 
and, in general, could result from dynamical relaxation of these
systems of massive protoplanets in several hundred thousand years
(Terquem \& Papaloizou 2002; Papaloizou \& Terquem 2001). 
The gravitationally bound masses of the planets remaining at the end
of the simulations  (calculated by considering all particles for 
which $2P + U < 0$, where $P$ is the pressure and $U$ is the gravitational 
binding energy) range between $2.4$
and $6.6 M_J$ in the higher disk mass case, and between $0.07$ and $1.7 M_J$
in the lower disk mass case. Hence both super-Jupiters and planets with masses
as small as that of Saturn seem to be a possible outcome of the instability
mechanism. This is an important feature of the model --- its natural 
flexibility in accounting for the entire range of masses of gas giants known so far.

The final masses of the planets are not just the result of merging
but also of accretion of ambient gas, and the accretion rate
in turn depends on the ambient pressure and thus on the equation
of state of the gas (D'Angelo, Kley \& Hennings 2003).
In the extended simulations the equation of state is
normally adiabatic, which means high pressure support of the surrounding gas 
(no cooling), and thus should yield a lower limit on the accretion rate. The mean accretion rate 
measured over the 500 years following the onset of fragmentation in the
run employing model DISL1  is 
quite low, in the range $10^{-7}-10^{-6} M_{\odot}/yr$ (different accretion rates are found for the different planets, in particular
the lowest are found for the planet closest to the star, as expected
from the higher ambient pressure). The accretion rate is declining
towards the end of the simulations, and during the last couple of
orbital times is practically zero. Therefore, at least in these 
adiabatic runs the measured protoplanetary masses after $\sim 1000$ years are
certainly a good estimate of the final ones.
The mass accretion rate onto the central star, instead, is still high even
at the end of the simulations, being $> 10^{-6} M_{\odot}/yr$
 (see Figure 16); using the value of the accretion 
rate in the last stage of the simulation would yield a disk dissipation 
timescale as small as $20000$ years (this is likely a lower limit given that 
the accretion  rate is still declining at the end).

In passing we note that the short disk dispersal timescales predicted
here as a result of gravitational instability would solve
the puzzle of short disk lifetimes (shorter than a million year in at
least 30\% of stars in Taurus, see Armitage, Clarke \& Palla 2003). 
Other solutions, like photoevaporative
flows (Clarke, Gendrin \& Sotomayor 2001; Armitage, Clarke \& Palla 2003)
require an input from external irradiation sources since the heat generated
by internal star-disk accretion shocks is probably insufficient (Matsuyama,
Johnstone \& Hartmann 2002), but this, for example, would not work in 
Taurus because there are no massive stars capable of generating such a strong 
photoevaporating flux. One might worry that the accretion rate of gas
towards the center, and, in general, any motion of the particles in the
disk, might be partly caused by the artificial viscosity, which is well known
to produce spurious losses of angular momentum (Thacker et al. 2000).
We tested this latter possibility by stopping mass accretion in model DISgr
when $M_d = 0.01 M_{\odot}$ (the mass is an order of magnitude
smaller than our typical disk masses) and running it forward in time
for 20 orbital times. The disk remains very smooth in this case ($Q_{min}
> 8$), and the accretion rate towards the end is nearly two 
orders of magnitude lower than in the other runs, 
being stationary at around $3 \times 10^{-7} M_{\odot}/yr$. We interpret the last number as the residual accretion rate due to artificial viscosity (note that this is a conservative choice because the disk
is never perfectly axisymmetric due to the inevitable  Poisson noise
in the initial conditions); the corresponding accretion timescale (defined
as the time required by a massive disk of about $0.1 M_{\odot}$ to accrete
onto the star by the latter mechanism only) is 
close to half a million year, so much longer than any of the timescales 
considered here. Hence the fact that a high accretion rate is
a necessary consequence of gravitational 
instability and provides  a way to clear out the disk very rapidly 
seems a well grounded inference. 
Indeed, Pickett et al. (2003) have recently obtained similar accretion
rates (although they probed the disk evolution on a shorter timescale)
using high resolution grid-based simulations that have a much lower
numerical viscosity. 
Notwithstanding the high mass of the protoplanets, most of the mass at the end
of the simulations, about $70\%$, is still in the disks; of this mass almost
$30\%$ is accumulated in the inner 2-3 AU as a result of accretion triggered 
by the non-axisymmetric torques and, partly, by artificial viscosity, 
while the rest is still at $R > 5$ AU, and hence will be still important 
in determining the
orbital evolution (and eventual migration) of the planets (see Lufkin et 
al. 2003). Accretion rates peak at values higher than $10^{-5} M_{\odot}/yr$,
significantly in excess of the rate measured in T Tauri stars
(Gullbring et al. 1998); 
the strong bursts of infrared luminosity found 
in disks caught during the early stages of their 
evolution, like FU Orionis, are suggestive of such high accretion rates and might 
indeed be explained with such strong inflows due to gravitational instability
(Lodato \& Bertin 2003).

We tested how accretion rates  depend on the
equation of state by re-simulating for a few orbital times model DISL1 with
a locally isothermal equation of state during and past fragmentation. 
The protoplanetary masses are four times 
higher than in the adiabatic simulation after 600 years (at this point 
we stop the simulation), the time-averaged accretion rate being about 50 times 
higher.
By this time, however, planets have also carved gaps that are not present in the adiabatic runs (compare Figure 15 with last snapshot of Figure 14); 
although material
can still flow to the planet, the accretion rate is strongly reduced
after gap formation (Bryden et al. 1999), in particular is about
a hundred times smaller
than the rate at which gas accretes onto the central object; therefore the
disk will be dissipated well before the planets can grow 
significantly further (Figure 16).
Once planets have carved a gap, further accretion of disk material
should occur on the viscous timescale; the only physical source of viscosity
in our disks is self-gravity (Lin \& Pringle 1987; Laughlin, Korchagin \&
Adams 1997) because, by design, we lack other possible sources like, for example,magnetic fields. The effect of artificial viscosity (Bryden et al. 1999), 
on the other end, is negligible at our resolution, as we discussed above.
Therefore, at this point the accretion rate onto the protoplanets should    in 
principle be proportional to that onto the central star.
However, not
everywhere in the disk is the dominant bulk motion directed towards
the center.
In fact, the direction of the torques arising from gravitational instability 
changes with radius, and, typically, while the material in the inner regions
loses angular momentum, the material in the outer part gains it.
The transition radius is typically set by where the dominant unstable
modes (the spiral arms in our case) occur, which in our simulations is
always between 10 and 15 AU from the center; therefore we expect that the
protoplanets located at $R< 10$ AU should accrete much more mass and much
more rapidly than those located at $R > 10$ AU (the actual accretion
rate will be determined by the local flux of mass around the
protoplanet, hence by the local details of the gravitational
torques).
The final masses of the protoplanets in this last run are between 10 and 25 $M_J$, therefore intermediate between those of brown dwarfs and those
of extrasolar planets (Udry et al. 2002).
However, we believe that these numbers are not to
be taken seriously. Indeed, 
by marking particles belonging to a clump at the final time
and tracing them back we found that most of the material accreted by the 
protoplanets  comes from the midplane and usually from a narrow annulus
coplanar with the orbit of the planet in the adiabatic runs, while it 
occurs in a much more isotropic fashion in the isothermal runs, with
a large fraction of accreted particles originally located at high
distances from the plane; this big difference is certainly due to the fact
that the vertical pressure gradients and those across the spiral shocks
near the planet are artificially low in the isothermal simulation
(shock heating is instantaneously damped). In fact, after
600 years of evolution the vertical structure of the isothermal and adiabatic
run are dramatically different, the scale height being almost ten times as big
in the latter. This is reflected in the comparison between the vertical
temperature profiles (Figure 17).

In general,  a conclusive answer on the final masses of the 
planets formed via gravitational instability
has to await a more realistic treatment of the disk thermodynamics,
with heating and cooling correctly model both inside and outside the
overdense regions (Mayer et al., in preparation).  

There are hints that protoplanets undergo some orbital migration 
(Mayer et al. 2002), but this does not seem to have a preferred direction 
due to the chaotic nature of the torques present in the
system (as a result of both the non-axisymmetric global potential and 
the gravitational interactions with the other planets). This supports the idea that current models of planet 
migration become unrealistic once non-trivial disk profiles (Artimowicz \& Peplinski, in preparation) and interactions
between several bodies are simultaneously taken into account 
(Lufkin et al. 2003).
Although the protoplanets are massive, the ratio of their masses to that
of the disk left after nearly a thousand years of evolution is still as
small as a few times $10^{-2}$ in the adiabatic run; therefore (Ward 1997a,b;
Tanaka et al. 2002) we do not 
expect migration to occur on the viscous timescale of the disk as in the classic 
Type II regime (gap formation is not apparent indeed) instead gravitational 
torques should 
be still the dominant source like in the Type I regime, albeit with the additional 
complication of disk self gravity and mutual interaction between the protoplanets. 
On the contrary, in the isothermal run the protoplanets acquire much higher masses due to the higher
accretion rates, gaps are carved by them and the orbital
evolution resembles that of  the Type II regime (see above) with the difference that
the disk is being dissipated very quickly.
Whereas we are not able to probe the systems for 
timescales long enough to draw conclusions, an efficient inward 
orbital migration is hard to imagine in the traditional
framework because the disks acquire a significantly steeper
inner density gradient (inside the orbits of the planets) due to 
accretion of gas onto the central star ---  outward migration should thus 
be more likely (Masset \& Papaloizou 2003). 

So does this mean that hot
Jupiters are difficult to explain within the present model? 
Not necessarily. There are indeed several possibilities to have 
efficient migration as soon as we move out of the standard framework.
Indeed, one of the mechanisms recently proposed to
explain the observed orbital distribution of extrasolar planets relies
on dynamical relaxation of a population of massive planets formed rapidly
through gravitational instabilities (Papaloizou \& Terquem  2001). This 
latter model assumes that the disk is dissipated on a timescale much shorter 
than that required for relaxation (the latter being
of order of thousand of orbits, or several tens of thousand years for the
orbital timescales typical of our simulations), an assumption that
seems to be marginally supported by our simulations.
As mentioned above, when the disk is still present, the
net gravitational torque caused by the instability will have different
directions depending on the location in the disk.
In the locally isothermal run with model DISL1
there is one protoplanet significantly inside 10 AU, and this might sink
towards the center in only a few thousand years if
it just follows the ``bulk'' accretion flow estimated 
for the disk (Figure 16).
The same would happen even in the adiabatic run, but
of course the sinking timescale will be longer this time, of order of a few
tens of thousand years. Therefore, one could
speculate that if protoplanets can migrate inwards just 
enough to find themselves inside the region where the torques due to 
gravitational instability become negative then they could
drift towards the center rapidly together with the rest of the disk,
eventually stopping at some distance from the star only after  
the non-axisymmetric torques have faded away; of course how much the 
planet can sink thanks to this latter mechanism is completely
undetermined at the moment, but future simulations are on the way
that will probe much longer timescales and possibly provide the answer.

Aside from the latter possibilities, 
certainly a more straightforward prediction of our model is
that inward migration should not be very efficient; this is actually a
good feature of disk instability --- that of planets sinking too fast 
towards the central star has become an increasingly hard problem
to solve in the past few years within the framework of the
standard core-accretion model (e.g. Bate et al. 2003).
Ways that have been proposed to halt
migration, like the interaction between a planet on an eccentric orbit
and the surrounding disk (Papaloizou \& Larwood 2000; Papaloizou 2002)
are naturally included in the gravitational instability model studied here,
although the individual effects are hard to disentangle given the complexity
of the evolution. 
On the other hand, we have to keep in mind that our simulations still lack 
several ingredients like stellar winds, photoionization and magnetic turbulence
(Balbus \& Hawley 1991; Nelson et al. 2003; Matsuyama, Johnstone \& Murray, 2003) that 
can affect substantially the inner disk, eventually creating cavities or 
severely affecting the density profile, with consequent effects on the speed
and direction of migration.

The surviving protoplanets are differentially rotating, 
nearly spherical bodies slightly flattened by
rotation (the ratio between major and minor axis is $\sim 0.9$ - 
see Figure 18). 
We measure equatorial rotation speeds 
(this is measured at the outermost radius for which particles are 
gravitationally bound to the clump) and then calculate a tentative final 
rotation speed of the protoplanets by allowing contraction down to the mean
density of Jupiter assuming conservation of angular momentum. We find values
in the range 3.5-40 km/s, which nicely encompass the (equatorial) rotation 
speeds measured for Saturn and Jupiter, 8.7 and 14.6 km/s respectively.
We note that a long standing problem has always been how Jupiter can still
maintain a high rotation speed despite the fact that some kinetic energy 
must have been dissipated by atmospheric friction after its formation; the 
present model suggests a solution in that protoplanets could form 
with rotation speeds well in excess of the speed that  Jupiter has today.
Among the surviving protoplanets the obliquities go from a few degrees to 160
degrees.  Large obliquities seem to be associated with mergers (and thus
transfers of angular momentum); the planet
experiencing fewer mergers is in both runs that with the smaller obliquity.
Our model would then naturally explain the wide range of obliquities that we 
find in our Solar System once we consider gas and ice giants together 
(a possible common origin of gas and ice giant planets via gravitational 
instability is discussed in Boss 2002b and 
Boss, Wetherill \& Haghighipour 2002).

\section{Conclusions}

The main result shown in this paper is that fragmentation into long-lived,
tidally stable, gravitationally bound protoplanets with masses and orbits
comparable with those of observed extrasolar planets is possible in marginally
unstable protoplanetary disks ($Q_{min} < 1.4$). The requirement is that
the disk can cool efficiently (as implicit in the locally isothermal 
approximation) until the spiral arms approach fragmentation; once the
the local density grows by roughly an order of magnitude gravity is strong
enough for the collapse to proceed even with purely adiabatic conditions.
We also showed that resolution, both in mass and in the gravitational
force calculation, is
a decisive factor in order to model disk evolution and fragmentation properly.
In  SPH simulations ,in particular, it is crucial that gravity and 
pressure be resolved at comparable levels for most of the extent of
the simulations. In fact, no matter how many particles are used, if the 
gravitational softening is too large, the spiral arms do not reach
the critical amplitude for fragmentation as the dynamical response of 
the system is altered --- growing modes are suppressed.
Once fragmentation is approached, a high resolution is also needed for the 
clumps to continue collapsing. Because the survival of clumps subject to
strong tidal stresses depends on their binding energy, thus on the
density they are able to reach, the fact that in our simulations clumps
survive for several tens of orbital times 
is also a consequence of the high resolution employed. 
Their densities are several order of
magnitudes higher than the mean density, resulting in a tidal radius
ten times larger than their typical size. Therefore they will eventually 
survive for
timescales much longer than those probed here and can thus be
associated with protoplanetary objects. Their destruction can only result
through mergers with other protoplanets or from a strongly increased 
tidal field in case their orbit migrates inward substantially. 
Pressure gradients near the clumps might drive dust and planetesimals
and enrich the gaseous protoplanets up to metallicities beyond the
solar value (Haghighipour \& Boss 2003).

Previous works on the gravitational instability fell short of the
resolution needed to follow the very non-linear stage of disk evolution.
In addition, fixed boundaries were certainly a problem; in run DISL1, for
example, both the outermost and the innermost clump go, respectively, further
out and further in than the initial outer and inner radius of the disk,
due to their eccentric orbits; therefore, even with enough resolution 
typical fixed grids would have not been able to follow 2 out of 3 clumps (see
also Pickett et al. 2000a and Boss 2000). Recently, Pickett et al. (2003)
identified several ``banana-shaped'' overdensities in their grid simulations;
these structures have densities and shapes strikingly similar to what
we find in our disks just before clump formation starts.
As they discuss and test, their simulations seem to lack enough azimuthal
resolution to be conclusive about the evolution of the overdensities.
As a comparison their grid cell size is $\sim 5$ times bigger than the
gravitational softening in our $10^6$ particles runs in the outer,
more unstable regions; indeed in runs where
the gravitational softening is increased by a factor of 3 or more with
respect to the nominal value (DISL1b,c in Table 1) we also witness a 
suppression of fragmentation  (section 3.3).

Whereas the global non-axisymmetric instabilities seen in the simulations
and the resulting fragmentation cannot be captured by the WKB approximation,
the maximum scales of the overdensities in the mildly non-linear regime
seem to be understandable in terms of the
maximum Toomre wavelength. The minimum masses of the forming clumps are
instead controlled by the local Jeans mass; because of the scaling with 
disk mass and temperature, disks with similar $Q$ profiles can produce
smaller or bigger clumps depending on their mass and temperature.
The smallest clumps
have masses lower than that of Saturn at formation. Further mass growth due
to merging and accretion shifts the typical masses to objects comparable
or bigger than Jupiter, but still it is clear that  gravitational
instability does not produce only super-Jupiters. Indeed, the mean
mass among the surviving protoplanets in the extended runs is $\sim 2.5
M_j$.

We  investigated whether starting a disk with a low $Q_{min}$ at
the beginning of a simulation, as we always do, might artificially enhance fragmentation. 
We showed that a disk grown from a very light state over tens of
orbital times still produces several gravitationally bound protoplanets
once it reaches values of temperature and mass comparable to the
initial ones of a model that undergoes clump formation (section 3.5).
This latter result
as well as the fact that disks with $Q_{min}$ just above the threshold
for fragmentation have a $Q$ profile similar to the initial one after
the  instability saturates (Figure 1), suggest that the non-axisymmetric 
instabilities occurring when $Q_{min} > 1.4$ are too weak to redistribute 
as much mass as needed to suppress the instability;a little cooling 
will easily bring the disk towards a marginally unstable state 
with $Q_{min} < 1.4$. 
Therefore $Q_{min} \sim 1.4$ appears to be a significant threshold. In reality
the growth of the disk mass with time will be determined by the balance 
between the accretion rate onto the central star as determined by both 
gravitational instability and other processes, for example viscosity
produced by magnetic fields, and the accretion rate of material falling 
onto the disk
from the molecular cloud envelope. Therefore the mass will not grow
at a constant rate as assumed here, instead the process will be strongly
time dependent; however, hydrodynamical simulations of disk formation
(Yorke \& Bodenheimer 1999), that include radiative transfer but neglect
magnetic fields, do find that disks reach $Q$ parameters in the range $1.3-1.5$
early in their evolution.
Whether  fragmentation will
actually occur will then depend on how well a disk can radiate away the 
thermal energy  produced by compression and shocks along the edges of 
the growing spiral arms. This is the most important, still open question 
concerning the final outcome of gravitational instabilities with the 
inclusion of realistic
thermodynamics (Pickett et al. 1998,2000a,b, 2003; Meija et al. 2003). 
In fact, even in our growing disk simulation we 
were keeping the local temperature constant before the appearance of the
overdensities. Recent, lower resolution SPH simulations that 
solve for heating and cooling find that fragmentation can proceed when 
the cooling time is comparable to the disk orbital time (Rice et al. 2003),
basically confirming previous simpler numerical and analytic calculations
by Gammie (2001). Similar conclusions are reached in the recent 
three-dimensional calculations with volumetric cooling by Pickett et al. 
(2003). Such short cooling timescales are actually achieved in the simulations of Boss 
(2001, 2002a,b), that include radiative transfer in the diffusion 
approximation with realistic disk opacities, due to efficient vertical energy
transport by convection.
We will address these issues in a forthcoming paper using very high resolution
simulations which incorporate different forms of radiative cooling. These
simulations will also allow to model more realistically the accretion rate
onto the protoplanets and thus produce a better prediction for their
masses. 

Finally, within the gravitational instability model the appearance
of the protoplanets and a rapid disk dispersal seem to be linked; although
it is premature to estimate a robust disk dispersal timescale (even
this can vary depending on the way the disk thermodynamics is treated), our
calculations suggest that most of the disk material originally at tens of
AU from the central star will be  accreted in less than $10^5$ years. Material 
originally located outside the strongly unstable region, say at distances 
of about 100 AU, would gain the angular momentum shed  by the 
strong spiral arms and avoid rapid accretion;
therefore a prediction of the gravitational instability model is that
there must be a population of fairly young protoplanetary disks (considerably
less than a million years old) in which a gap in the mass density of gas
exists over a few tens of AU between an inner and an outer zone.
SIRTF and other upcoming missions (Evans et al. 2003), by looking in 
the mid-infrared wavelengths,
will allow for the first time to trace the structure and evolution of 
the gaseous component in the protoplanetary disks using the rotational 
emission lines of its main constituent, molecular hydrogen,
and will provide a direct estimate of disk dispersal timescales.

\bigskip

L.M. thanks all the participants to the workshop ``Circumstellar disks
and protoplanets'' organized by Tristan Guillot at the Nice Observatory
in February 2003, in particular Hal Levinson, Doug Lin, Pawel Artimowicz, 
Tristan Guillot, Paolo Tanga, Patrick Michel, Alessandro Morbidelli and Ricardo Hueso for 
insightful and stimulating discussions. The numerical simulations were 
performed on LeMieux at the Pittsburgh Supercomputing Center, on the Z-Box
at the University of Zurich and on the Intel cluster at the
Cineca Supercomputing Center in Bologna (Italy).

\begin{figure}
\plotone{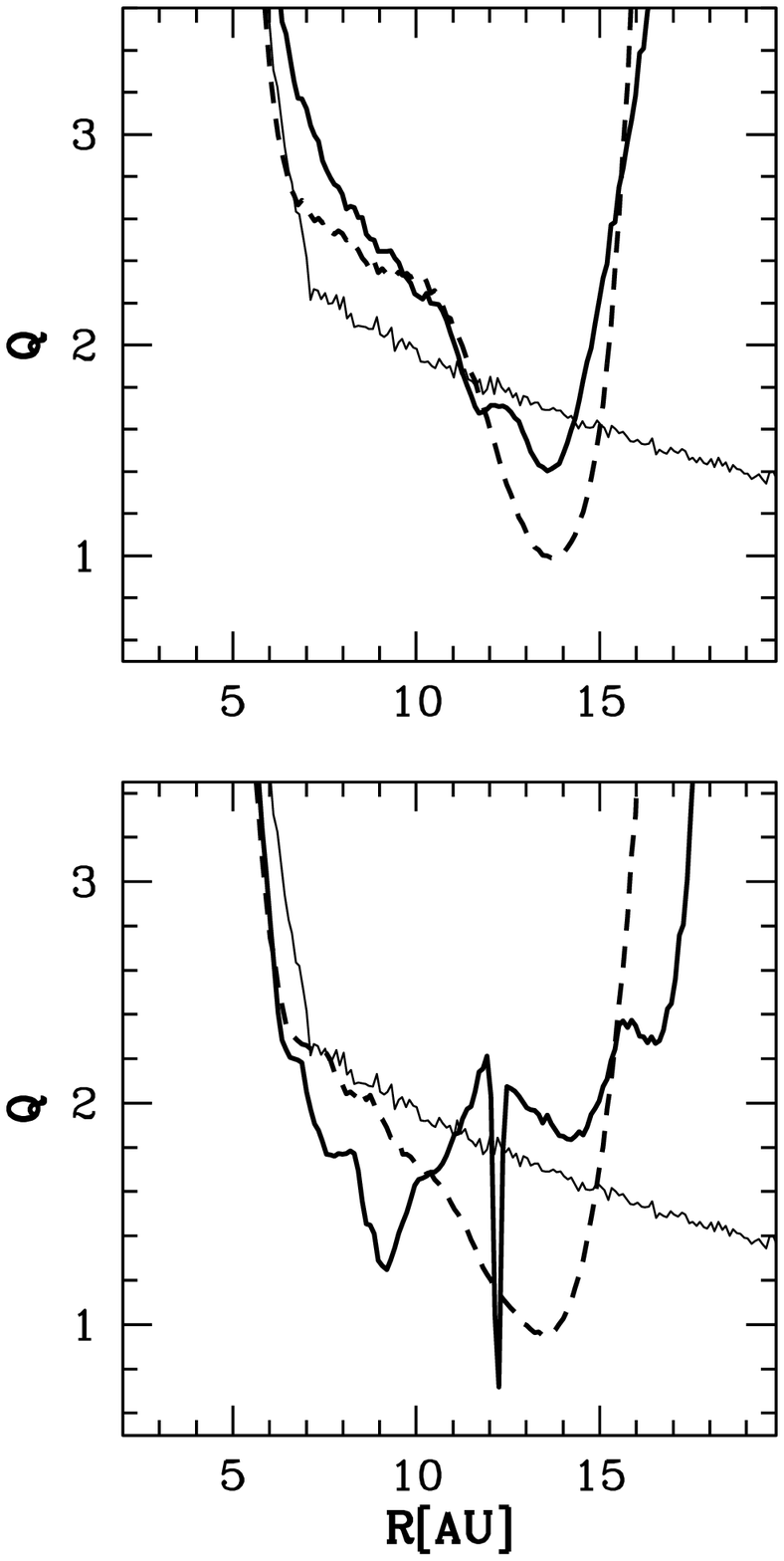}
\caption{Evolution of Q profiles. Upper panel: model DISH2. Lower panel:model
DISH1. We show the profiles at t=0 (thin solid line), t=160 years (dashed line) and t=240 years (thick solid line). Fragmentation occurs between 160 years and
240 years in model DISH1, while model DISH2 develops only strong spiral
arms.}
\end{figure}

\begin{figure}
\plotone{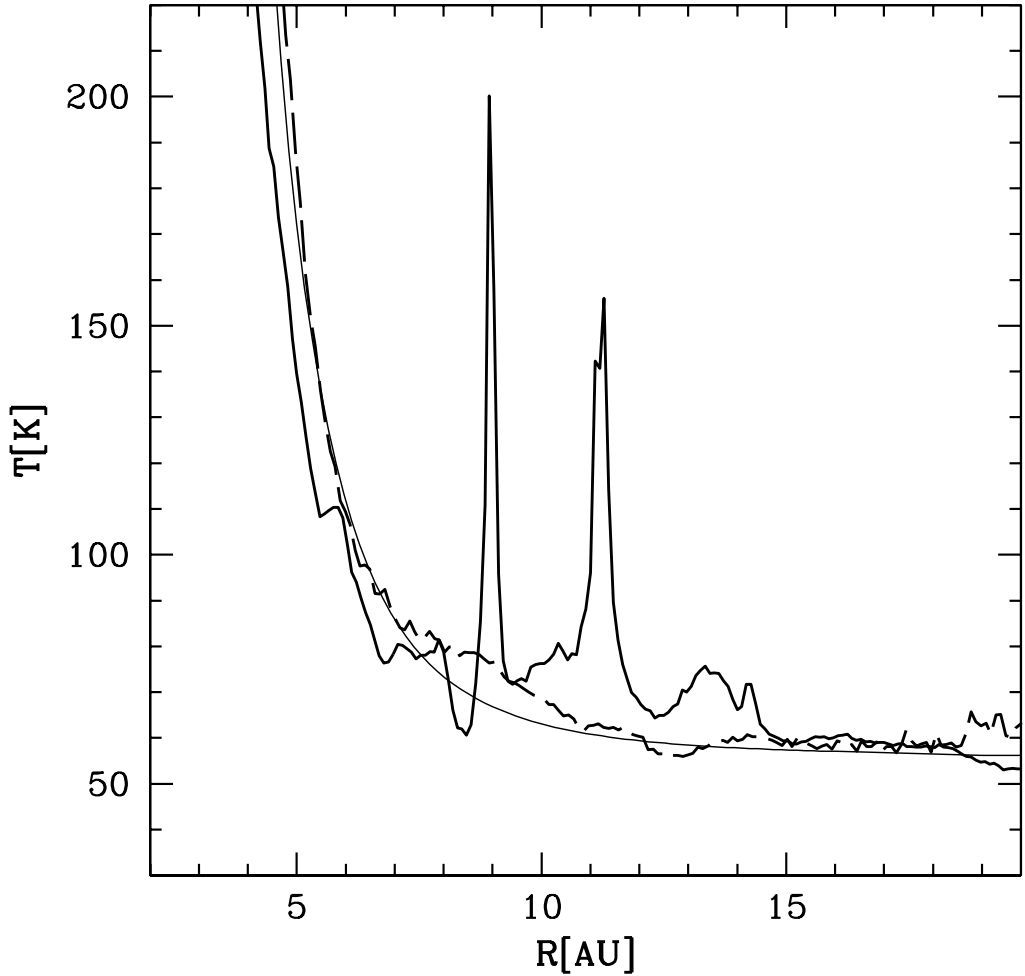}
\caption{Evolution of the temperature profile. The initial profile 
(thin solid line) for an outer temperature of 56 K is shown --- this profile
was used for both model DISH1 and DISH2 (see Table 1). Model DISH1 undergoes
clump formation and its profile is shown at T=320 years (thick solid line), 
after the equation of state has been switched to adiabatic (see text). 
The peaks correspond to regions where bound clumps are 
(a single peak contains more than one clump due to limited bin size).
Model DISH2 only forms spiral arms; its profile (thick dashed line)
is also shown at T=320 years 
(the equation of state, in this case, is locally isothermal throughout the evolution, so the 
small changes of the temperature with radius are due to radial diffusion
of particles only).}
\end{figure}

\begin{figure}
\plotone{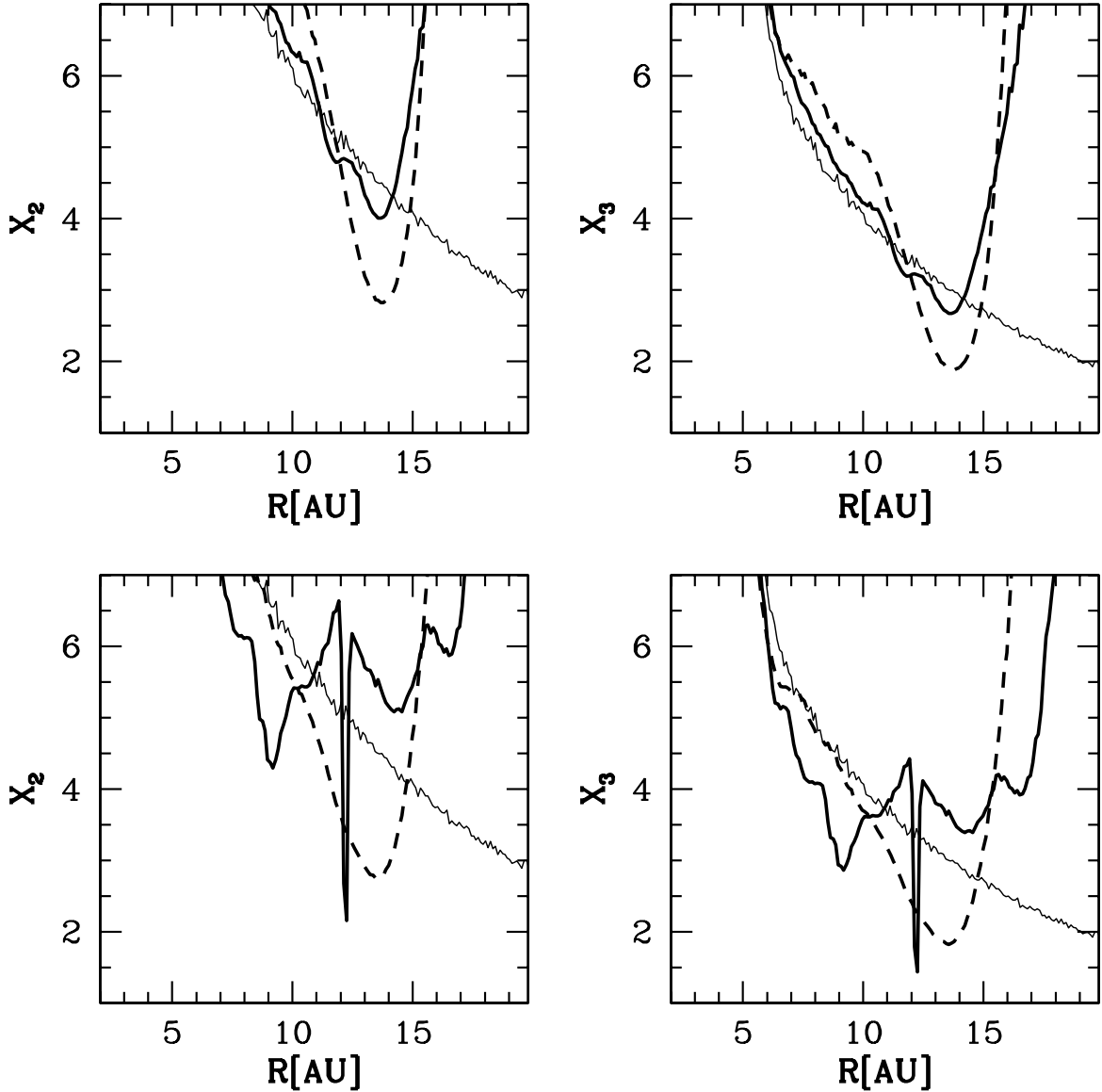}
\caption{Evolution of $X_{2}$ (left) and $X_3$ (right) profiles. 
Upper panels: model DISH2. Lower panels:model DISH1. Profiles are shown at 
t=0 (thin solid line), t=160 years (dashed line) and t=240 years (thick solid
line). See also Figure 1 on the Q profiles for the same models.}
\end{figure}

\begin{figure}
\plotone{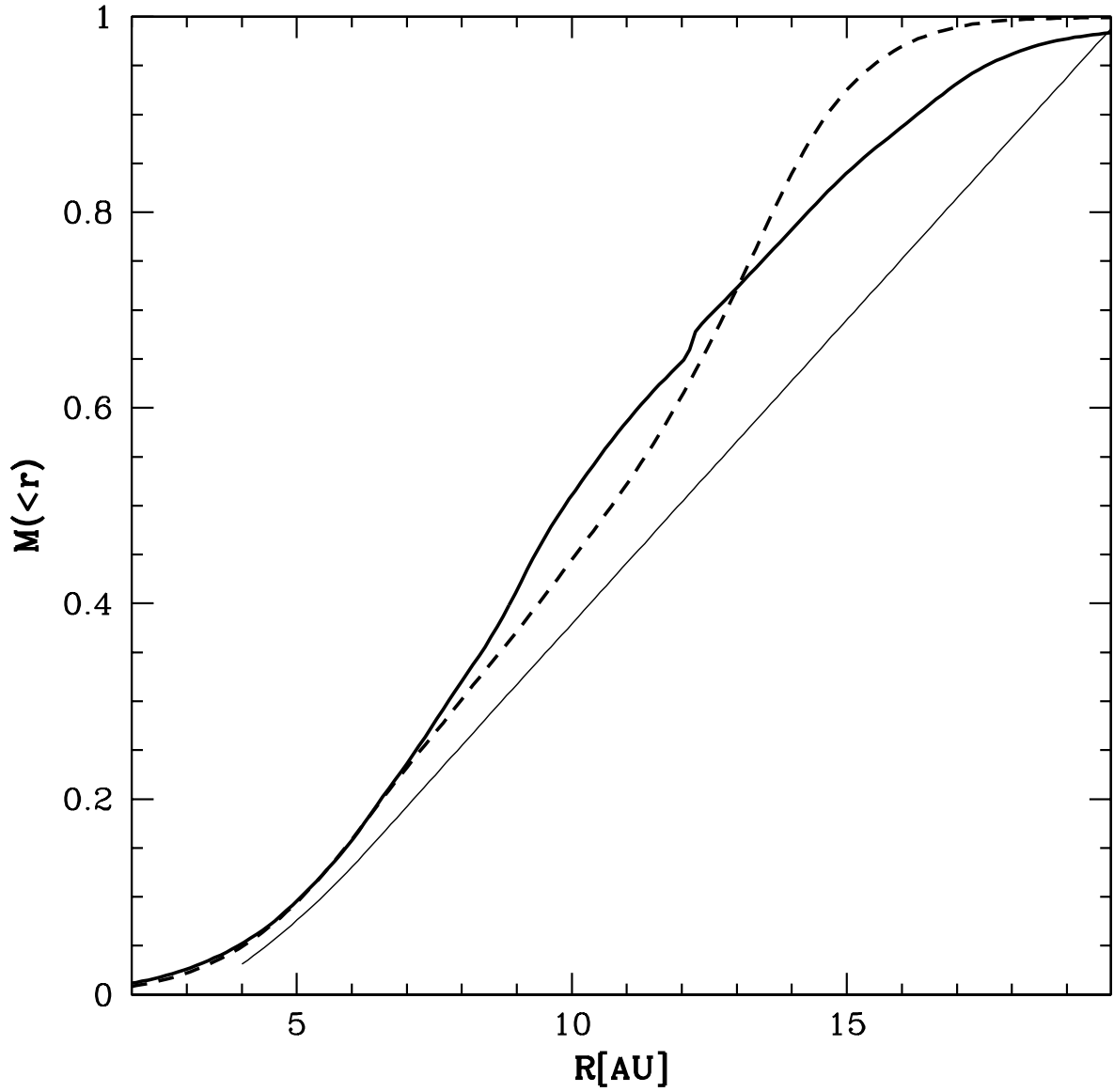}
\caption{Evolution of the cumulative surface mass profile of model DISH1. The
profile at T=0 (thin solid line), at T=200 years (dashed line - at this time
the spiral arms have the strongest amplitude and are about to fragment) and
at T=350 years (thick solid line - fragmentation has already occurred).}
\end{figure}

\begin{figure}
\caption{Color-coded density maps of run DISH1 (see Table 1)
after 350 years when the locally isothermal equation of state 
is used throughout the 
entire calculation (left panel) and when the simulation switches to an 
adiabatic equation of state (right panel) once
the overdensities have grown past some threshold (see text, section 3.1)
Brighter colors are for higher densities (densities between $10^{-14}$ and
$10^{-6}$ g/cm$^3$ are shown using a logarithmic scale - the same applies to
all density maps shown in this paper) and the disks are shown out
to 20 AU.}
\end{figure}

\begin{figure}
\caption{Color-coded face-on density maps of run DISH3 (upper panels, $Q_{min} \sim 1.3$)
and DISH3b (lower panels, $Q_{min} \sim 1.5$), at 200 (left) and 
350 years (right).
See Table 1 for details on the models. The equation of state is
switched to adiabatic close to fragmentation in run DISH3.
Brighter colors are for higher densities (see Figure 5) and the disks 
are shown out to 20 AU.}
\end{figure}

\begin{figure}
\plotone{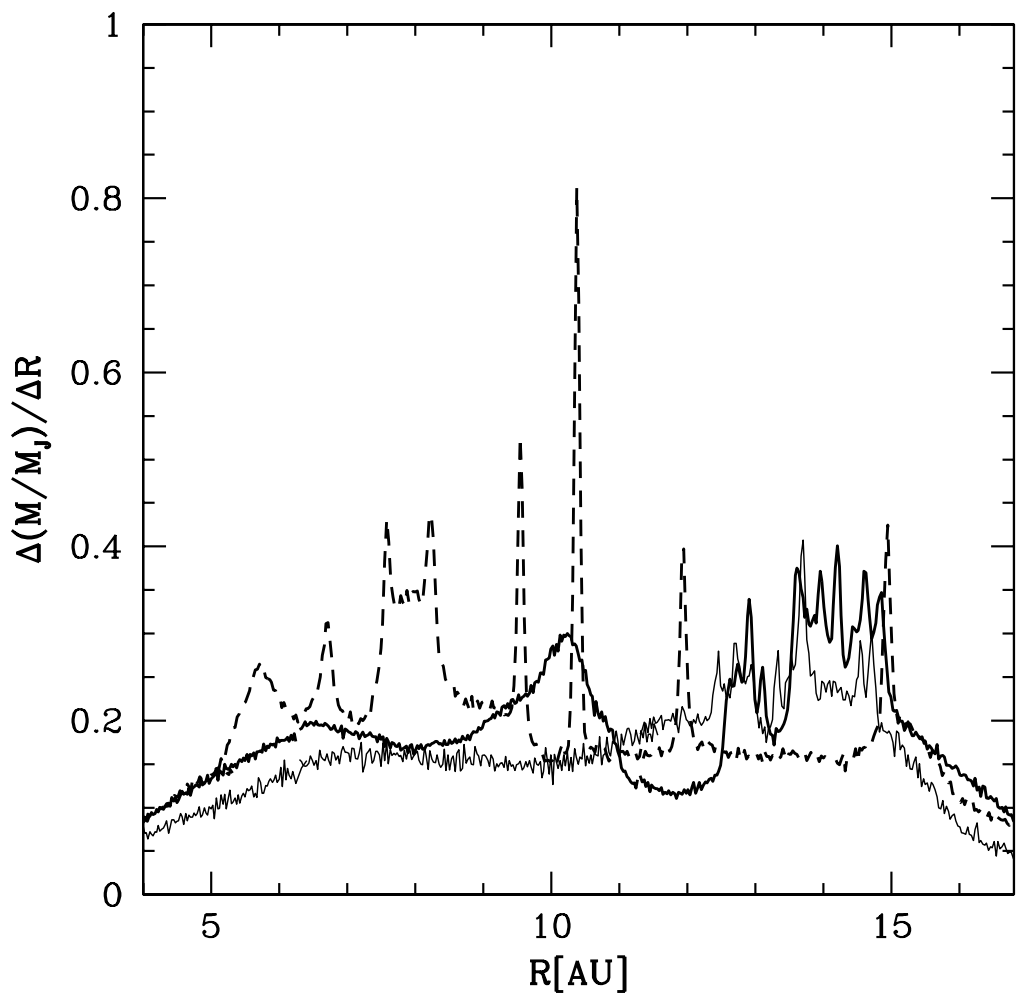}
\caption{Radial mass profile at fragmentation. Results are shown for model 
DISH1 (dashed line), model DISH3 (thick solid line) and model DISgr (thin
solid line). The mass is measured in units of one Jupiter mass, $M_J$,
using cylindrical bins equally spaced in radius.
The peaks correspond to bound clumps. Although the
difference in the  masses of the disks is small (e.g. DISH1
versus DISH3) the difference in the clumping mass scale is large because
of the way Jeans length scales with mass and temperature 
(see section 3.1)}
\end{figure}

\begin{figure}
\caption{Color-coded face-on density maps of run DISL4 after 120 (left
panel) and 240 years (right panel), out to a radius if 25 AU. Note 
the stronger spiral arms compared to the runs shown in Figure 5, and 
the very eccentric orbital trail of the outermost clump in the right panel.}
\end{figure}

\begin{figure}
\plotone{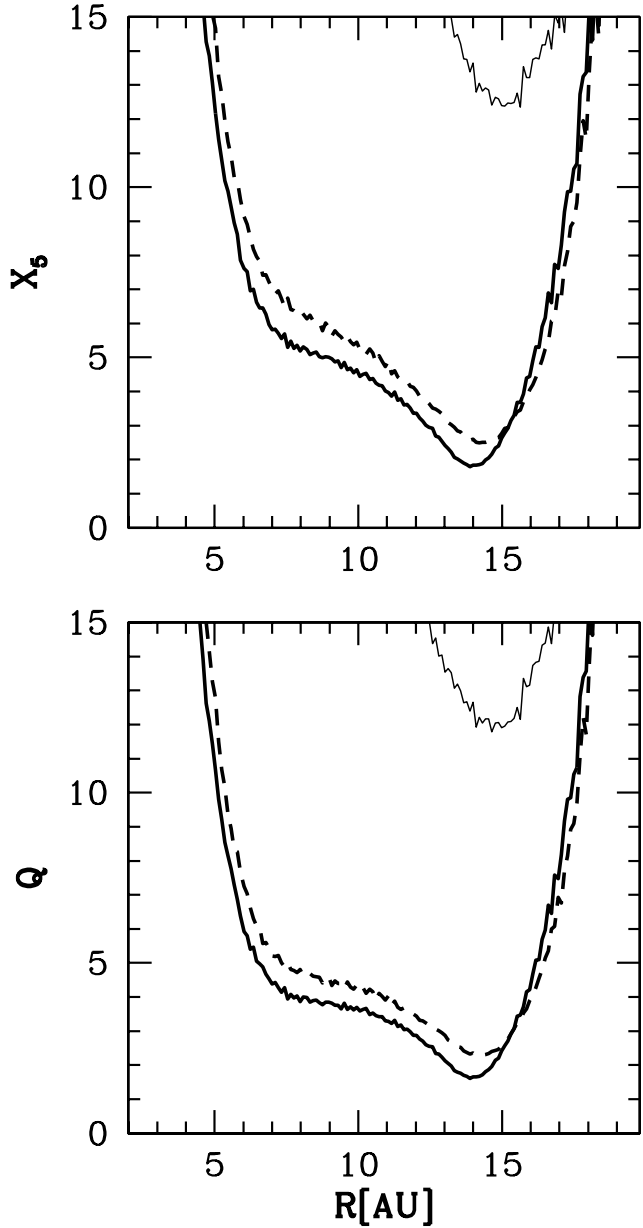}
\caption{Evolution of the Toomre Q parameter (bottom) and of the 
$X_5$ parameter (top) for the growing disk in run
DISgr (see Table and section 3.1.1.). The profiles is shown at 300 years 
(thin solid line), 450 years (dashed line) and 500 years 
(thick solid line - just before fragmentation starts. Note that the $X_5$
parameter is just below the threshold for instability ($<3$) even 
at late times, which might explain why high order modes dominate 
in this run.}
\end{figure}

\begin{figure}
\caption{Color-coded face-on density maps of adiabatic runs out to 20 AU.
Brighter colors are used for higher densities (see Figure 1).
The two upper panels show model DISLad3 at,
respectively, 200 (left) and 300 (right) years, 
while the two lower panels show model DISLad4
at, at 200 (left) and 300 (right) years, respectively.}
\end{figure}

\begin{figure}
\caption{Color-coded face-on temperature maps of the adiabatic run DISLad4. 
Brighter colors are used for higher temperature (the scale goes from 20 to 2000 K). Two snapshots are shown at, respectively, 120 years and 220 years (at this
point clumps have just started forming), out to 20 AU from the central star.
Note the strong increase of temperature along the edges of the spiral arms 
and at the location of the clumps.}
\end{figure}

\begin{figure}
\plotone{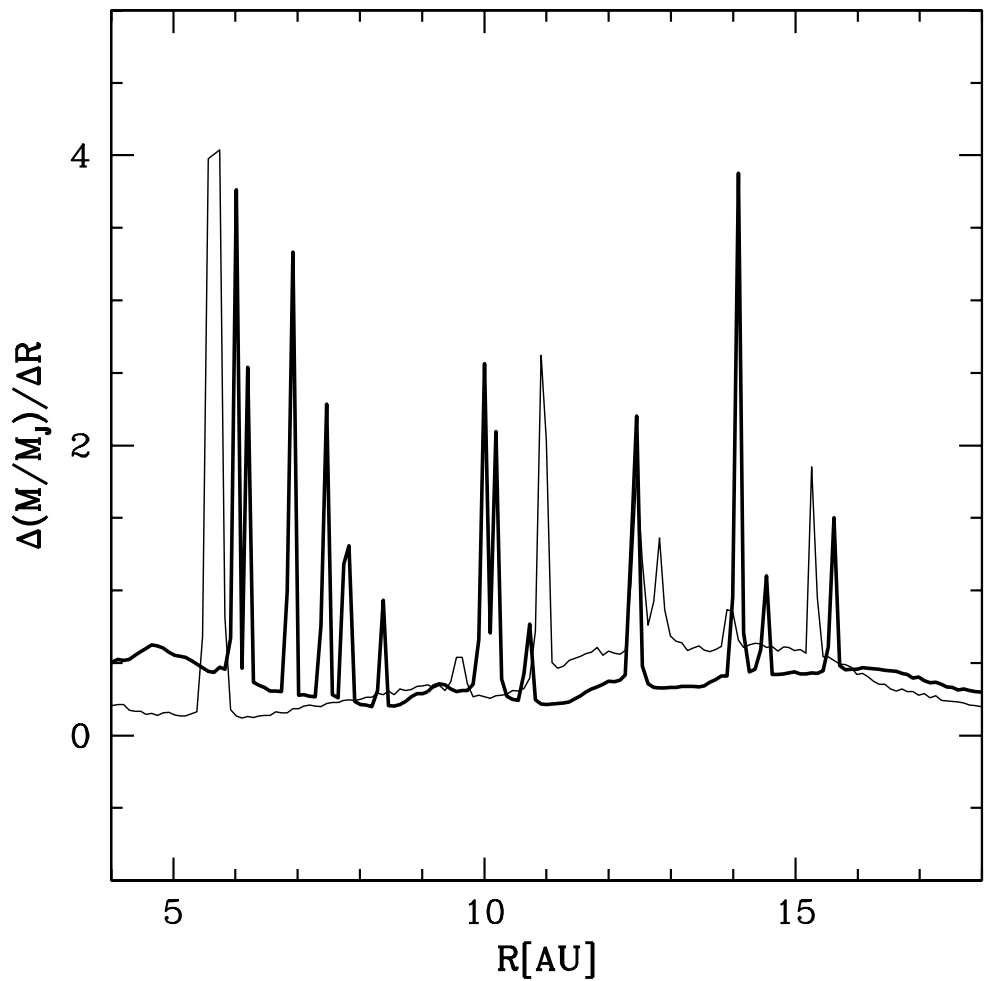}
\caption{Differential mass profile after 350 years in two 
simulations with same disk model and different resolutions, run DISH1 (thick solid
line) and run DISL1(thin solid line). The mass is measured in cylindrical radial bins,
the unit is 1 Jupiter mass, $M_J$. See Table 1 for details on the simulations.
Peaks correspond to gravitationally bound clumps.
Clearly several more clumps are present in the higher resolution 
run.}
\end{figure}

\begin{figure}
\plotone{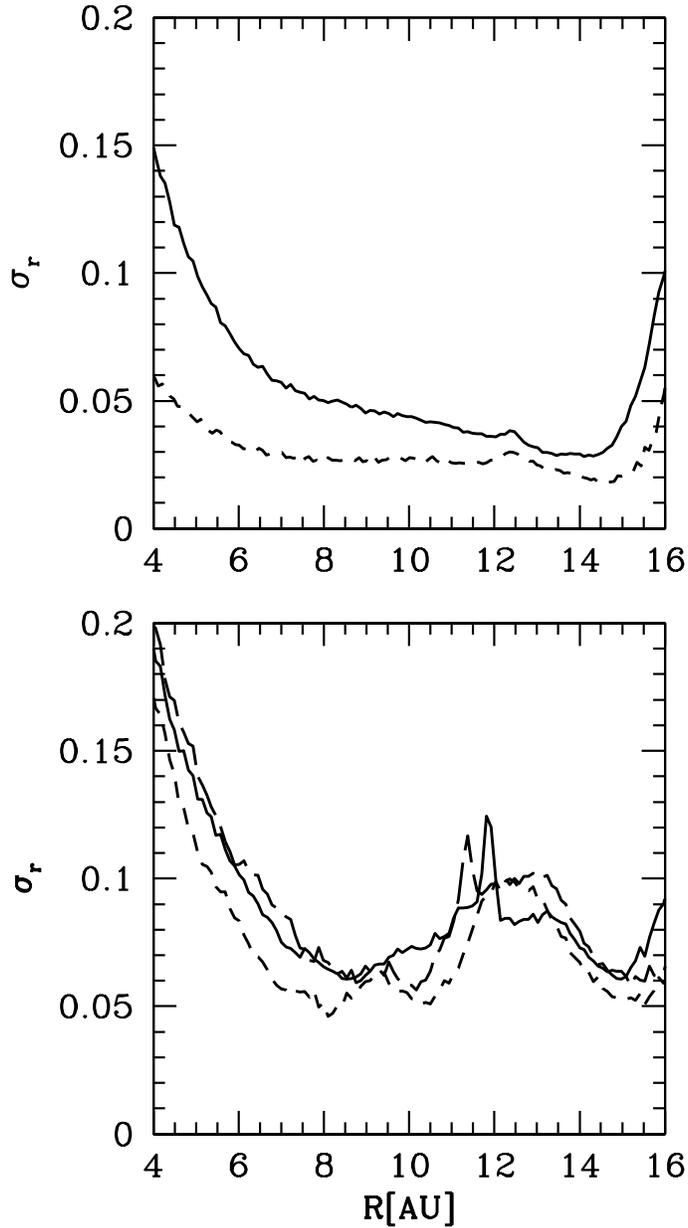}
\caption{Effect of varying artificial viscosity on the velocity field of 
the disk in locally isothermal simulations. The radial velocity dispersion as a 
function of radius is shown. Regions where the radial dispersion is higher 
correspond to zones of higher non-axisymmetric motion or even collapse (the
peaks).
On top two weakly unstable models, DISH2 (solid line) and DISH2b 
(dashed line) are shown after  200 years of evolution; at the bottom 
three strongly unstable models, DISL1 (solid line), DISL1e (long-dashed line) 
and DISL1f (short-dashed line) are shown at the time of maximum growth of 
the spiral overdensities (just before fragmentation in the case
of run DISL1, after about 160 years.)}
\end{figure}

\begin{figure}
\caption{Color-coded density maps of the growing disk simulation, run DISgr.
Brighter colors are for higher densities (see Figure 1) and the disk is 
shown out to 20 AU. From top to bottom and clockwise, snapshots are 
taken after  300, 480, 560 and 800 years.}
\end{figure}

\begin{figure}
\caption{Color-coded face-on density map of a run employing model DISL1 with
a locally isothermal equation of state even after the appearance of the
overdensities (see text, section 3.5). The box is 30 AU on a side and the snapshot is
taken after 450 years. Compared to the simulations employing
adiabatic conditions once the overdensity threshold is reached
(see Figure 14 and Figure 2 in Mayer et al. (2002)), it is evident that 
protoplanets are carving much clearer gaps, considerably more mass
is piling up at the center and the disk is being dispersed much more
quickly (the mean density is very low except in the central regions).}
\end{figure}

\begin{figure}
\plotone{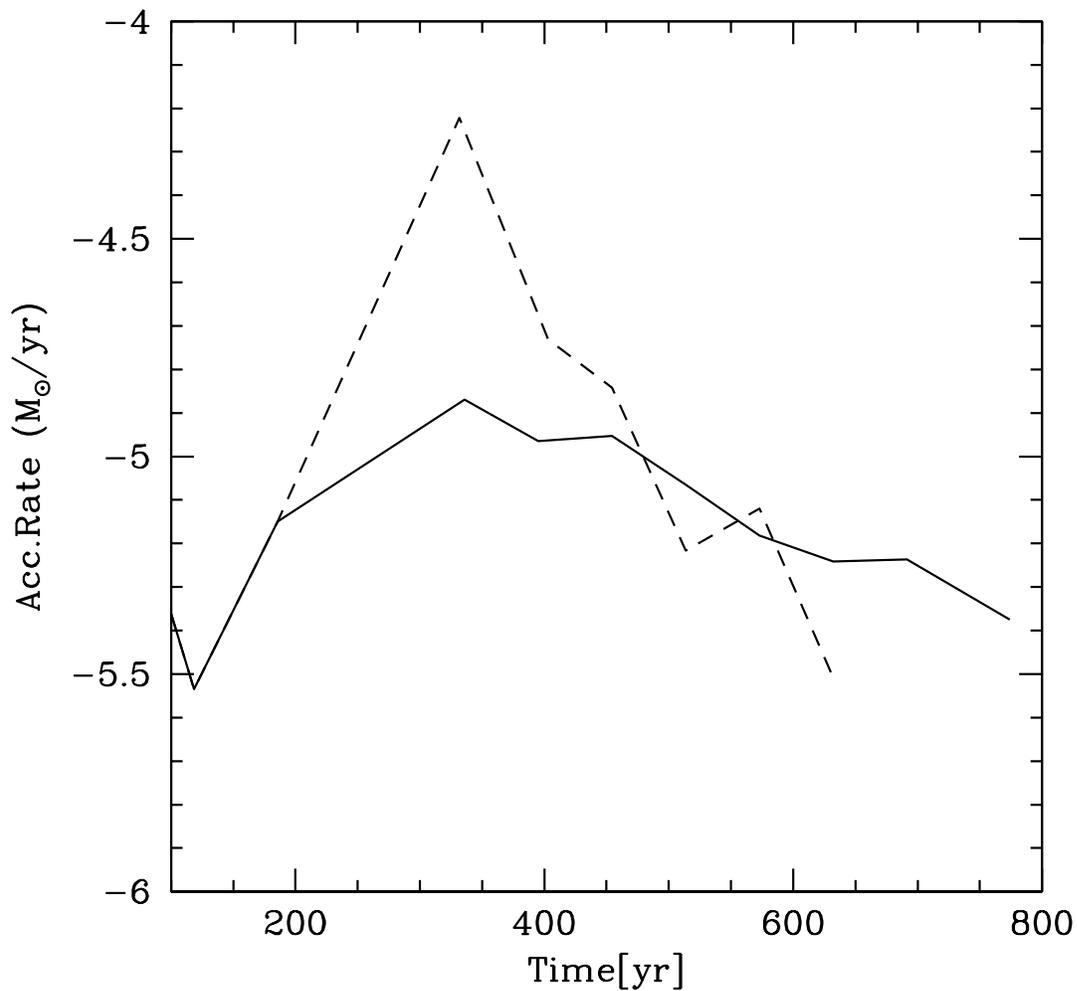}
\caption{Accretion rate of gas onto the central star. The flow of mass inside
a region of size equivalent to the gravitational softening length of
the star (2 AU) is calculated. The simulations employ model DISL1 (see
Table 1), the solid line is used for the case in which the equation
of state is switched to adiabatic above the assigned density
threshold, while the dashed line refers to the case in which the equation
of state is locally isothermal for the entire duration of the calculation
(see text, section 3.5).}
\end{figure}

\begin{figure}
\plotone{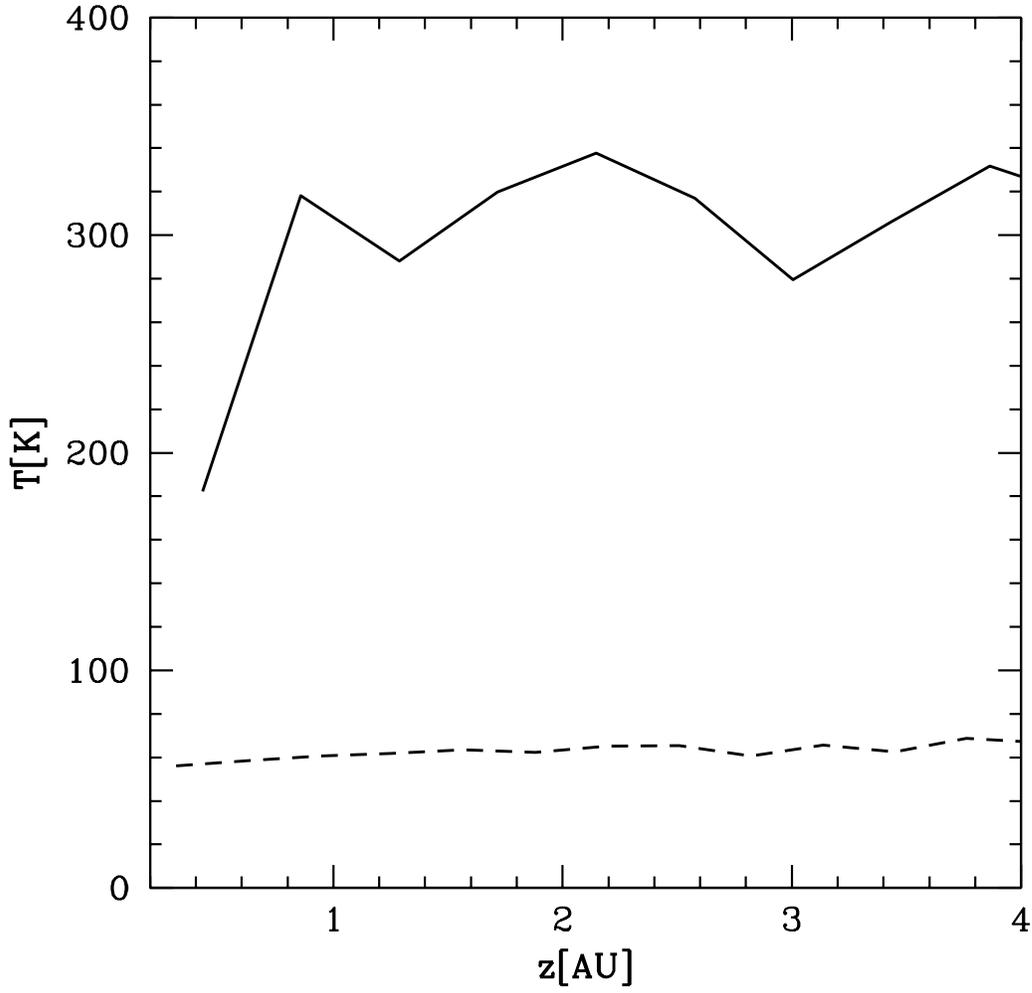}
\caption{Temperature profile along the disk vertical axis for model
DISL1 after 600 years for a fixed locally isothermal equation
of state (dashed line)
and for the case in which the equation of state is switched
to adiabatic prior to clump formation (solid line).}
\end{figure}

\begin{figure}
\caption{Close-up view of a gravitationally bound clump in run
DISH1 (with the adiabatic switch) at 350 years. We show the projection 
perpendicular to the angular momentum axis of the clump.
A color-coded plot of the velocity field is shown,
the redder the color the higher the velocity. The clump is clearly
in differential rotation. The box extends out to 0.5 AU and contains
about 17574 particles.}
\end{figure}

\clearpage

\begin{table}
\begin{center}
\caption{Parameters of the simulations.
Column 2: Initial disk mass as a fraction of stellar mass. Column 3: 
Equation of state (ISO for isothermal, otherwise $\gamma$ is indicated). 
Column 4: $\alpha$ parameter of
artificial viscosity; Column 5: $\beta$ parameter in artificial
viscosity. Column 7: number of disk particles
Column 8: Softening of disk particles (in AU)
Column 9: Mass of central star (in $M_{\odot}$); Column 10: Toomre Q parameter;
Column 11: Outcome of the simulations in terms of clump formation 
("yes" for formation of gravitationally bound clumps, 
"no" for no fragmentation at all, or ``transient'' for transient clumps).}
\begin{tabular}{lcccccccccc}
Run &  $M_d/M_s$ & $EOS$ & $\alpha$ & $\beta$ & $N_p$ & $T_{min}$ & softening & $M_s$ & $Q$ & clumps \\ 
\\
DISL1  & 0.1 & ISO & 1 & 2 & $2 \times 10^5$ & 56 & 0.12 & 1 & 1.38 & yes \\
DISL1b & 0.1 & ISO & 1 & 2 & $2 \times 10^5$ & 56 & 0.18 & 1& 1.38 & no \\
DISL1c & 0.1 & ISO & 1 & 2 & $2 \times 10^5$ & 56 &  0.6 & 1 & 1.38 & no \\
DISL1d    & 0.1 & ISO & 1 & 2 & $2 \times 10^5$ & 56  & 0.06 & 1 & 1.38 & yes\\
DISL1e  & 0.1 & ISO & 1 & 2.5 & $2 \times 10^5$ & 56 & 0.12 & 1 & 1.38 & yes\\
DISL1f  & 0.1 & ISO & 1 & 3 & $2 \times 10^5$ & 56 & 0.12  & 1 & 1.38 & no\\
DISL1g    & 0.1 & ISO & 1 & 6 & $2 \times 10^5$ & 56 & 0.12 & 1 & 1.38 & no\\
DISL1h      & 0.1 & ISO & 1 & 0.5 & $2 \times 10^5$ & 56 & 0.12 & 1 & 1.38 & yes\\
DISL1i   & 0.1 & ISO & 0 & 0.5 & $2 \times 10^5$ & 56 & 0.12 & 1 & 1.38 & yes \\
DISL2     & 0.1 & ISO & 1 & 2 & $2 \times 10^5$ & 100 & 0.06 & 1 & 2 & no \\
DISH1       & 0.1 & ISO & 1 & 2 & $10^6$ & 56 & 0.06 & 1 & 1.38 & yes \\
DISH2   & 0.08 & ISO & 1 & 2 & $10^6$ & 56 & 0.06& 1 & 1.65 & no\\
DISH2b  &  0.08 & ISO & 0 & 0.5 & $10^6$ & 56 & 0.06 & 1 & 1.65 & no\\
DISH2c          & 0.08 & ISO & 0 & 0.5 & $10^6$ & 56 & 0.006 & 1 & 1.65 & yes \\
DISH3      & 0.085 & ISO & 1 & 2 & $10^6$ & 36 & 0.06 & 1 &  1.3 & yes \\
DISH3b      & 0.085 & ISO & 1 & 2 & $10^6$ & 50 & 0.06 & 1 & 1.5 & no \\ 
DISgr        & 0.0085 & ISO & 1 & 2 & $2 \times 10^5$ & 30(gr) & 0.06 & 1 & Q(t) & yes \\
DISH4     & 0.075 & ISO & 1 & 2 & $10^6$ & 56 & 0.06 & 1 & 1.9 & no \\
DISH4b    & 0.075 & ISO & 1 & 2 & $10^6$  & 56 & 0.006 & 1 & 1.9 & no \\
DISL3      & 0.075 & ISO & 1 & 2 & $2 \times 10^5$ & 56 & 0.06 & 1 & 1.9 & no \\
DISL4        & 0.075 & ISO & 1 & 2 & $2 \times 10^5$ & 56 & 0.06 & 0.5 & 1.38 & yes \\      
DISLad1      & 0.1 & 1.4 & 1 & 2 & $2 \times 10^5$ & 56 & 0.06 & 1 & 1.38 & no \\
DISLad2      & 0.125 & 1.4 & 1 & 2 & $2 \times 10^5$ & 20 & 0.06 & 1 & 0.8 & no \\
DISLad3      & 0.125 & 1.3 & 1 & 2 & $2 \times 10^5$ & 20 & 0.06 & 1 & 0.8 & transient \\
DISLad4      & 0.125 & 1.2 & 1 & 2 & $2 \times 10^5$ & 20 & 0.06 & 1 & 0.8 & yes \\
DISLad5      & 0.1 & 1.4 & 1 & 0.5 & $2 \times 10^5$ & 56 & 0.06 & 1 & 1.38 & no \\
DISLad6      & 0.1 & 1.4 & 0 & 0.5 & $2 \times 10^5$ & 56 & 0.06 & 1 & 1.38 & no \\

\end{tabular}
\label{t:simul}
\end{center}
\end{table}

\end{document}